\documentclass[prl,twocolumn,superscriptaddress,floatfix,noshowpacs,10pt,longbibliography]{revtex4-2}%

\usepackage{graphicx,bm,times}
\usepackage{amsmath}
\usepackage{amsfonts}
\usepackage{amssymb}
\usepackage{comment}
\usepackage{bm}% bold math
\usepackage{color}
\usepackage{xcolor}
\usepackage{times}
\usepackage{soul}
\usepackage[table]{xcolor}

\usepackage[%
colorlinks=true,
urlcolor=blue,
linkcolor=blue,
citecolor=blue
]{hyperref}
\definecolor{Orchid}{RGB}{218,112,214}

\begin{document}

\title{Field-induced incipient  spin-density  phase stabilized inside the nematic phase of 
FeSe$_{1-x}$S$_x$}

\author{I. Paulescu}
\email[corresponding author:]{ioana.paulescu@physics.ox.ac.uk}
\affiliation{Clarendon Laboratory, Department of Physics,
University of Oxford, Parks Road, Oxford OX1 3PU, UK}

\author{R. M. Abedin}
\affiliation{Clarendon Laboratory, Department of Physics,
University of Oxford, Parks Road, Oxford OX1 3PU, UK}

\author{J. S. Pearce}
\affiliation{Clarendon Laboratory, Department of Physics,
University of Oxford, Parks Road, Oxford OX1 3PU, UK}

\author{W. H. Fong}
\affiliation{Clarendon Laboratory, Department of Physics,
University of Oxford, Parks Road, Oxford OX1 3PU, UK}

\author{Z. Zajicek}
\affiliation{Clarendon Laboratory, Department of Physics,
	University of Oxford, Parks Road, Oxford OX1 3PU, UK}

\author{A. Morfoot}
\affiliation{Clarendon Laboratory, Department of Physics,
University of Oxford, Parks Road, Oxford OX1 3PU, UK}

\author{W. Knafo}
\affiliation{Laboratoire National des Champs Magnetiques Intenses, EMFL, CNRS, Univ. Grenoble Alpes, INSA-T, Univ. Toulouse 3, 31400 Toulouse, France}

\author{O. Squire}
\affiliation{Clarendon Laboratory, Department of Physics,
University of Oxford, Parks Road, Oxford OX1 3PU, UK}

\author{D. Graf}
\affiliation{National High Magnetic Field Laboratory and Department of Physics, Florida State University, Tallahassee, Florida 32306, USA}

% \author{D. Vignolles}
% \affiliation{Laboratoire National des Champs Magnetiques Intenses, EMFL, CNRS, Univ. Grenoble Alpes, INSA-T, Univ. Toulouse 3, 31400 Toulouse, France}

\author{A. A. Haghighirad}
\affiliation{Institute for Quantum Materials and Technologies, Karlsruhe Institute of Technology, Kaiserstr. 12, 76131 Karlsruhe, Germany}

\author{A. I. Coldea}
\email[corresponding author: ]{amalia.coldea@physics.ox.ac.uk}
\affiliation{Clarendon Laboratory, Department of Physics,
University of Oxford, Parks Road, Oxford OX1 3PU, UK}

\begin{abstract}
Spin-density wave (SDW) order and superconductivity frequently compete and coexist in unconventional superconductors, where spin fluctuations often mediate superconducting pairing \cite{Keimer2015, Shi2025, Maple1995,Fernandes2022}. In iron-chalcogenide superconductors, FeSe$_{1-x}$S$_x$, SDW order has only been detected under applied pressure, while both spin and nematic 
fluctuations are involved in determining their rich superconducting phase diagrams \cite{Sprau2016,Nag2025}. Here, we 
report evidence for an incipient SDW phase,
within the nematic state of FeSe$_{1-x}$S$_x$, revealed in magnetic fields up to 68~T. Once superconductivity is quenched, we observe 
sharp upturns in longitudinal resistivity accompanied by anomalies in tunnel diode 
oscillator frequency response and torque anisotropy, consistent with a field-induced electronic order. Dominant low-frequency quantum 
oscillations reveal a small reconstructed Fermi surface, consistent with a field-induced SDW order. Direct experimental comparisons with a pressure-tuned nematic, analogue, FeSe$_{0.96}$S$_{0.04}$,  demonstrate that SDW phases are stabilized within the nematic phase of FeSe$_{1-x}$S$_x$ via both chemical substitution and applied pressure. These findings reveal
that by weakening nematicity,
the SDW orders are  stabilised, which
promotes  dominant superconducting pairing mechanism in iron chalcogenides.
\end{abstract}

\date{\today}
\maketitle

{\bf INTRODUCTION.}

Iron-based superconductors display complex magnetic ground states that involve both itinerant and local magnetism \cite{Dai2012}, and are often characterized as Hund's metals \cite{Fernandes2022,deMedici2014,Dai2015}.
 In the iron pnictides, the parent compounds typically display stripe-type SDW order that is closely connected to quasi-nesting between hole and electron pockets,
  which naturally amplifies the spin susceptibility at particular wave vectors in the presence of interactions
  and large densities of states at the Fermi level \cite{Mazin2008,Terashima2009}. 
 Superconductivity generally appears when this magnetic order is weakened by doping or pressure \cite{Chu2009,Kimber2009}. These systems provide the canonical case of microscopic coexistence and competition \cite{Cvetkovic2009,Fernandes2010,Vorontsov2010}
 in which SDW order gaps selected portions of the Fermi surface and suppresses superconductivity, but the residual spin fluctuations near the SDW instability could promote a sign-changing order parameter \cite{Mazin2008,Hirschfeld2011}.

 By contrast, in iron chalcogenides, the relevant magnetic landscape is often more frustrated, with several nearly degenerate configurations competing over a narrow energy scale \cite{Fernandes2022,Glasbrenner2015}.
Among them, 
FeSe is highly unusual,
 as it harbors 
a nematic electronic state,
 which displays strong orbitally-dependent electronic anisotropies and interactions \cite{Amalia2018,Amalia2021},
in the absence of any long-range static magnetic order at ambient pressure \cite{Bohmer2016,Wiecki2018}. Neutron scattering studies on strained-detwinned FeSe suggest large local fluctuating moments with competing Néel and stripe configurations \cite{Wang2016,Liu2025}. 
Despite the absence of static order at ambient conditions in FeSe, several magnetic configurations remain close in energy and may therefore be stabilized by different external perturbations.

\begin{figure*}[htbp]
	\centering
	 \includegraphics[trim={0cm 0cm 0cm 0cm}, width=0.9\linewidth,clip=true]{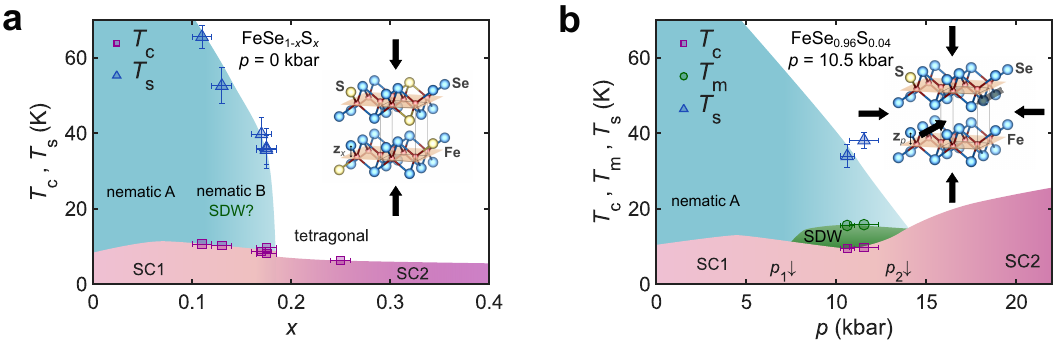}
	\caption{\textbf{Phase diagrams of FeSe$_{1-x}$S$_{x}$ tuned by chemical and applied pressure.} \textbf{(a)} Phase diagram, $T-x$, as a function of sulfur substitution, $x$ in the absence of magnetic field. Systems investigated in this study, $x=0.11-0.17(5)$ (blue triangles) lie close to the border of the nematic electronic phase indicated by $T_{\rm s}$ which is suppressed close to $x\sim0.18$. Superconducting regions with different pairing mechanisms, SC1 and SC2, emerge on either side of the nematic endpoint below $T_{\rm c}$ (pink squares). Schematic boundaries are constructed from refs. \cite{Bristow2020, Amalia2021,Bristow_thesis}
    and based on data in Supplementary Fig.~S1 and Supplementary Fig.~S2. 
    \textbf{(b)} Phase diagram of FeSe$_{0.96}$S$_{0.04}$ as a function of applied pressure in zero magnetic field (after Ref.~\cite{Zajicek2026}).
 Anomalies in resistivity suggest the
 presence of SDW inside the nematic phase
  (green region). The extracted temperatures also carry an estimated systematic error of the order $0.5$~K, due to differences between cooling and warming measurements. 
  The insets show the crystal structure of FeSe$_{1-x}$S$_{x}$ with atoms shown as circles for Fe (red), Se (blue) and S (yellow). The tetragonal unit cell is displayed by the black solid lines. The chalcogen height above the Fe plane is indicated by the parameter $z_x$ ($z_p$), tuned either by the chemical pressure in \textbf{(a)} or applied pressure in
  \textbf{(b)} and visualized by the black arrows. 
  }\label{FeSeS13pc_B_T_phase_diagram_zero_field}
\end{figure*}

 The isoelectronic FeSe$_{1-x}$S$_x$ are ideally suited to disentangle  
the role of different electronic phases competing with superconductivity using both chemical and applied
pressure (see Fig.~\ref{FeSeS13pc_B_T_phase_diagram_zero_field}) \cite{Amalia2021,Amalia2018,Bristow2020,Zajicek2026}.
The nematic endpoint close to $x=0.175(5)$ (see Fig.~\ref{FeSeS13pc_B_T_phase_diagram_zero_field}(a))
divides the superconducting phase of FeSe$_{1-x}$S$_x$  into two distinct pairing states
\cite{Hanaguri2018}:
one suggested to be mediated by spin fluctuation (SC1), giving rise to the isotropic sign-changing $s_{\pm}$-gap symmetry 
inside the nematic phase \cite{Sprau2016,Wiecki2018}; the other one inside the tetragonal phase
is proposed to have an anisotropic
and nodal order parameter (SC2) mediated by
nematic critical fluctuations 
\cite{Nag2025}. Interestingly,
upon application of hydrostatic pressure in FeSe, an SDW phase stabilizes once the nematic phase is significantly suppressed \cite{Sun2016,Matsuura2017}.
Beyond the nematic boundaries in the high-pressure phase, the superconductivity of FeSe increases dramatically towards 36~K \cite{Sun2016} and it may coexist with the high-pressure magnetism 
\cite{Bendele2012,Kothapalli2016}.

By combining both
chemical and applied pressure tuning, the boundaries of the phase diagram of FeSe$_{1-x}$S$_{x}$ are shifted, as the nematic and magnetic phases are suppressed with increasing $x$ (see Fig.~\ref{FeSeS13pc_B_T_phase_diagram_zero_field}(b)), whereas superconductivity at high pressures remains rather robust to different substitutions \cite{Matsuura2017,Reiss2024,Zajicek2022Cupressure,Zajicek2026}. 
Resistivity upturns  were previously used to identify the development of the SDW phase in zero-magnetic field in different FeSe bulk samples \cite{Sun2016,Xiang2017}
and thin flakes \cite{Xie2021}. 
With increasing $x$, the signature of the SDW phase  is  progressively
suppressed, as resistivity anomalies are still detected inside the nematic phase
of FeSe$_{0.96}$S$_{0.04}$, as shown in  
 Fig.~\ref{FeSeS13pc_B_T_phase_diagram_zero_field}(b)
\cite{Zajicek2026}.
However, SDW becomes washed out in the high-pressure magnetic phase
of FeSe$_{1-x}$S$_{x}$ \cite{Matsuura2017}, upon reducing thickness
in thin flakes of FeSe \cite{Xie2021} or 
by introducing strong impurity scattering by Cu doping \cite{Zajicek2022}.
These studies highlight the strong sensitivity of SDW phases to different tuning parameters, which in turn influences the superconducting pairing in different regimes.

Magnetic fields provide a useful tool
to suppress superconductivity and expose an underlying normal electronic order.
Magnetotransport studies of FeSe$_{1-x}$S$_{x}$ suggest two distinct regions inside the nematic phase
(nematic A for $x<0.1$ and nematic B for $0.1<x<0.18$, as shown in Fig.~\ref{FeSeS13pc_B_T_phase_diagram_zero_field}(b)), characterized by different resistivity slopes in high magnetic fields \cite{Bristow2020}.
Additionally, quantum oscillations indicate
significant changes inside the nematic B phase of FeSe$_{1-x}$S$_x$  \cite{Coldea2019,Reiss2020}.
Here,  we provide evidence that in the presence of high magnetic fields, a fragile SDW-like order is stabilized inside the nematic B phase of FeSe$_{1-x}$S$_x$  \cite{Amalia2021}.
Our experimental evidence was collected using magnetotransport, tunnel diode oscillator, torque, and quantum oscillations measurements to probe field-induced electronic order in magnetic fields up to 68~T. 
We provide a direct comparison of the field-induced electronic behaviour in two compositions of 
 FeSe$_{1-x}$S$_x$, near the nematic phase boundary, tuned by either chemical or applied pressure (see Fig.~\ref{FeSeS13pc_B_T_phase_diagram_zero_field}).

\begin{figure*}[htbp]
	\centering
	\includegraphics[trim={0cm 0cm 0cm 0cm}, width=1\linewidth,clip=true]{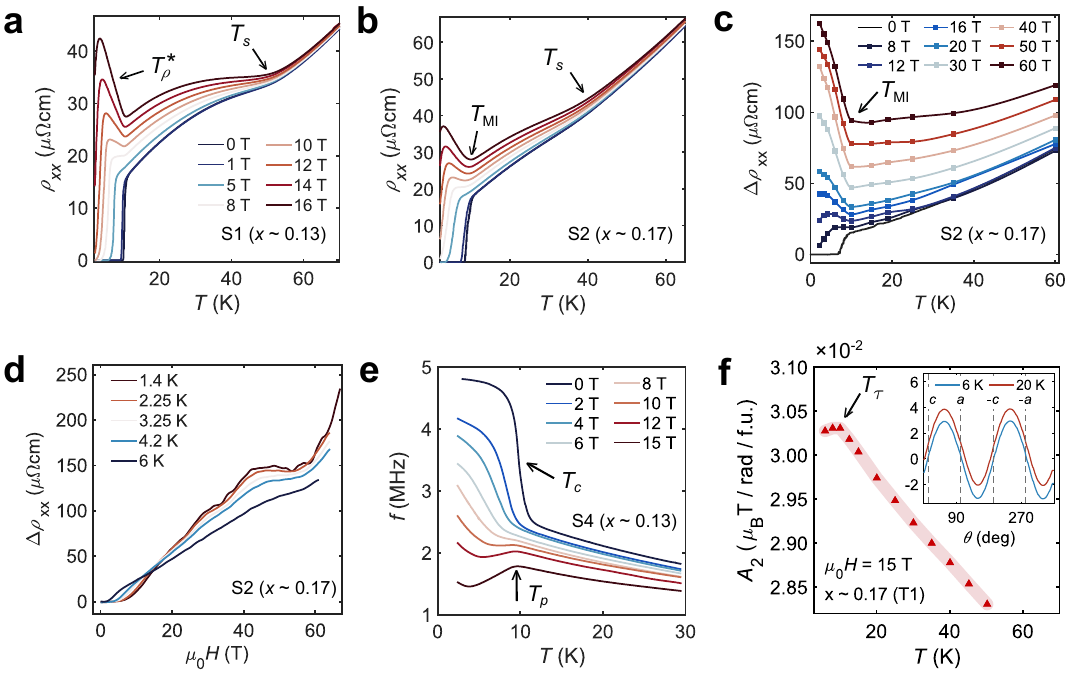}
	\caption{ \textbf{Transport and tunnel diode oscillator (TDO) behaviour of nematic FeSe\textsubscript{1-x}S\textsubscript{x} ($0.13<x<0.175$) in magnetic field.} \textbf{(a)-(b)} Longitudinal resistivity, $\rho_{xx}$, against temperature for two nematic samples S1 and S2 with different compositions, $x$, in magnetic fields up to 16\,T. $T_{\rm s}$ marks the nematic transition temperature of each sample. The emergence of resistivity upturns at low temperatures is quantified by the metallic-to insulating-like crossover at $T_{\rm MI}$ and
    additionally by the minimum in first-order derivative at $T^{*}_{\rho}$ (as defined in Supplementary Fig.~S5). \textbf{(c)} Resistivity change against temperature in zero field (solid black line) and at fixed magnetic fields (filled squares), as extracted from the high-field magnetoresistance curves of sample S2 shown in panel ({\bf d}). Solid lines connecting discrete data points are obtained from polynomial interpolation. \textbf{(d)} Changes in resistivity $\Delta\rho_{xx}$ as a function of magnetic field up to 68~T in different fixed temperatures. A constant value of $32.7\mu\Omega$cm was subtracted to account for an instrumental offset.
    A low frequency quantum oscillation dominates the signal below 6\,K (see Supplementary Fig.~S11).
      \textbf{(e)} Temperature dependence of the resonant TDO frequency, $f$, under different applied magnetic fields. The arrow marks the zero-field superconducting transition, $T_{\rm c}$ (maximum in the derivative shown in the Supplemental Fig.~S8), and a field-induced transition at the local maximum, $T_{p}$.
    \textbf{(f)} The temperature dependence of the two-fold FFT amplitude, $A_2$, of the magnetic torque for sample T1. The arrow at $T_\tau$ indicates the local maximum in the torque amplitude, and the solid red line is a guide to the eye. The inset shows the angular dependence of the magnetic torque at 6~K and 20~K in 15~T, by rotating the sample in the $(ac)$-plane, where traces are vertically offset for clarity. The dashed vertical lines signify the orientation of the applied field in relation to the crystallographic axes.}
	\label{Fig1_FeSeS_transport_TDO}
\end{figure*}

{\bf RESULTS.}

\vspace{0.2cm}
{\bf Resistivity and TDO resonant frequency response in magnetic fields.}
Figures~\ref{Fig1_FeSeS_transport_TDO}(a) and (b) show the evolution of longitudinal resistivity $\rho_{xx}$ against temperature with increasing applied magnetic field up to 16\,T for two different compositions of nematic FeSe$_{1-x}$S$_x$: sample S1~($x\sim0.13$ and
$T_{\rm s} \sim 52.4$\,K) and sample S2~($x\sim0.17$ and $T_{\rm s} \sim 39.5$\,K). In the absence of an applied magnetic field, both samples are superconducting below 10\,K, with large values of the residual resistivity ratio, ($RRR=\rho{(292~\rm K)}/\rho(T_{\rm on})>16$) and a small residual resistivity   ($\rho_{\rm xx}({T_{\rm on}}) \sim 16 \mu\Omega$cm, where $T_{\rm on}$ defines the onset of superconductivity as shown in Supplementary Fig. S4), 
in good agreement with previous results 
(see Supplementary Table~S1) \cite{Bristow2020}.
Additionally, the superconducting transition width, $\Delta T_{c}=T_{\rm on}-T_{\rm off}$, is less than 1\,K.
As superconductivity is suppressed by increasing the applied magnetic field above 5\,T, a pronounced upturn in resistivity develops at $T_{\rm MI}$, where the slope suddenly changes from metallic-like to insulating-like behavior (see Figure \ref{Fig1_FeSeS_transport_TDO}(b)).
Interestingly, with an increasing magnetic field
the position of $T_{\rm MI}$ shift towards lower temperatures, $\Delta T\sim 1$\,K over $\Delta B=5-16$\,T
(see Figure~\ref{Fig1_FeSeS_transport_TDO}(a)).
To follow this feature systematically, 
we notice similar trends for the field-dependence of $T^{*}_{\rho}$,  defined as the minimum in the first-order derivative of resistivity
as a function of temperature (see Supplementary Fig. S5), and previously
used to define the SDW anomaly in resistivity in BaFe$_2$As$_2$ \cite{Ikeda2018}.

\begin{figure*}[htbp]
	\centering
	    	\includegraphics[trim={0cm 0cm 0cm 0cm}, width=1\linewidth,clip=true]{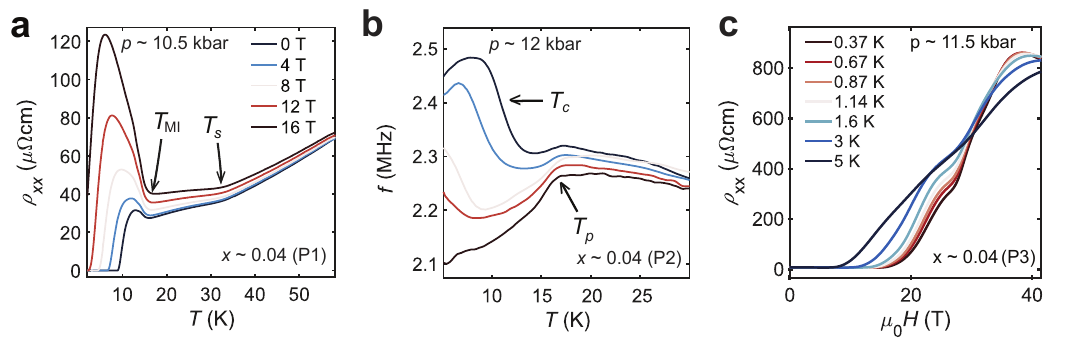}
	\caption{ \textbf{Temperature-dependent transport and tunnel diode oscillator (TDO) behaviour of nematic FeSe\textsubscript{0.96}S\textsubscript{0.04} under pressure in different magnetic fields.} \textbf{(a)} The longitudinal resistivity $\rho_{xx}$ against temperature under pressure of $p\sim 10.5$~kbar for sample P1 in fixed magnetic fields up to 16\,T aligned parallel to the $c$-axis. The arrow at $T_{\rm s}$ points out the nematic transition and $T_{\rm MI}$ indicates the metallic-insulating-like transition at the development of the SDW phase. \textbf{(b)} The temperature dependence of TDO resonant frequency, $f$,  in different magnetic fields. The arrows indicate the position of the magnetic transition at $T_p$ and the superconducting transition at $T_{\rm c}$. 
    \textbf{(c)} The field dependence of resistivity as a function of magnetic field up to 41~T in different fixed temperatures under pressure of 11.5~kbar for sample P3. A low frequency quantum oscillation dominates the signal below 5\,K (see Supplementary Fig.~S12). 
       }
    	\label{Fig2_FeSeS4pc_pressure}
\end{figure*}

To understand whether these anomalous upturns in resistivity are correlated with changes in electronic structure, longitudinal magnetoresistance was measured up to 68\,T at different constant temperatures for sample S2~($x\sim0.17$), as shown in Figure~\ref{Fig1_FeSeS_transport_TDO}(d). 
By slicing these isothermal field dependence data,
we construct the temperature dependence of resistance for sample S2 up to 60\,T
where the resistivity upturn is amplified in high fields to extremely large values (resistivity increases towards values measured in zero field around 100~K), as shown 
in Figure~\ref{Fig1_FeSeS_transport_TDO}(c). 
Additionally, in high magnetic fields, we detect a low-frequency quantum oscillation of $F\sim83$\,T, characteristic of a small Fermi surface pocket, with light effective mass
of $m^*$ = 1.53(5)~$m_e$
(see Supplementary Fig.~S11).
These findings are consistent with previous studies reporting the presence
of dominant slow oscillations in the nematic B phase of FeSe$_{1-x}$S$_x$ ($0.1<x<0.18$), which disappear in the tetragonal phase \cite{Coldea2019,Reiss2020}.

The tunnel diode oscillator (TDO) is a contactless technique sensitive to changes in both in-plane resistivity and susceptibility.
Figure~\ref{Fig1_FeSeS_transport_TDO}(e) shows the temperature dependence of the resonant TDO frequency with increasing magnetic fields of up to 15\,T for sample S4 ($x\sim0.13$). In the absence of a magnetic field, the TDO frequency has a clear increase  while entering the superconducting phase due to the diamagnetic response, leading to a drop in inductance ($T_{\rm c}$ is defined as the
maximum in the first derivative in Supplemental Fig.~S8
and listed in Supplemental Table~S1). 
As the magnetic field increases and suppresses superconductivity, a local maximum emerges at $T_{p}$ above 10\,T reflecting an increased skin depth due to changes in resistivity 
in the normal state
(see Figure~\ref{Fig1_FeSeS_transport_TDO}(e)). 
% The abrupt drop in TDO frequency in high magnetic fields is 
% , which can be confirmed by comparison with the resistivity in Figure~\ref{Fig1_FeSeS_transport_TDO}(a), 
These studies indicate a striking similarity in high magnetic fields between the resistivity upturn (see Figure~\ref{Fig1_FeSeS_transport_TDO}(a)) and the resonant frequency suppression (see Figure~\ref{Fig1_FeSeS_transport_TDO}(e)) (similar changes occur in $T^{*}_{p}$ and $T^{*}_{\rho}$ shown in Supplementary Fig.~S9 and Supplementary Fig.~S5).
Thus, the presence of field-induced anomalies in both TDO and transport reflects the development of a field-induced intrinsic electronic order,
similar to other systems, such as UTe$_2$\cite{Lin2020}.

Next, we employ magnetic torque to identify any changes in magnetic anisotropy in high magnetic fields.
 Here, the angular dependence of the torque has a sinusoidal dependence for polar-angle rotation  (see the inset of Figure~1(f)),
 probing the magnetic susceptibility anisotropy $\chi_a-\chi_c$. 
Taking into account previous sign conventions for stable equilibrium in torque \cite{Pearce2024}, the angular dependence of torque suggests $\chi_a>\chi_c$,  for FeSe$_{1-x}$S$_x$ ($x\sim0.17$), likely induced by the $g$-tensor anisotropy, induced by the spin-orbit coupling \cite{Pearce2024}.
The calculated torque is well below \(0.035\,\mu_\text{B}\text{T} \)/rad/f.u. at 15~T, confirming that the observed moment and anisotropy is intrinsically weak.
Interestingly, Figure~\ref{Fig1_FeSeS_transport_TDO}(f) shows that the corresponding two-fold amplitude, $A_{2}$, displays a local maximum at $T_{\rm \tau}$ in 15~T, close to the other field-induced anomalies observed in transport and TDO studies. Additionally, $A_{2}$ displays a change in slope at the nematic transition, $T_{\rm s}$, since the symmetry is reduced from tetragonal to orthorhombic phase, as evident in Supplementary Fig.~S10.
Previously,
 a strong anomaly in torque amplitude was identified at the SDW transition in BaFeAs$_{1-x}$P$_x$, coinciding with the
 anomaly in resistivity \cite{Kasahara2012}. Thus, the weak anomaly in the torque amplitude at $T_{\tau}$ is consistent 
 with the development of magnetic anisotropy
due to an incipient magnetic order.

\begin{figure*}[htbp]
	\centering
	 \includegraphics[trim={0cm 0cm 0cm 0cm}, width=0.9\linewidth,clip=true]{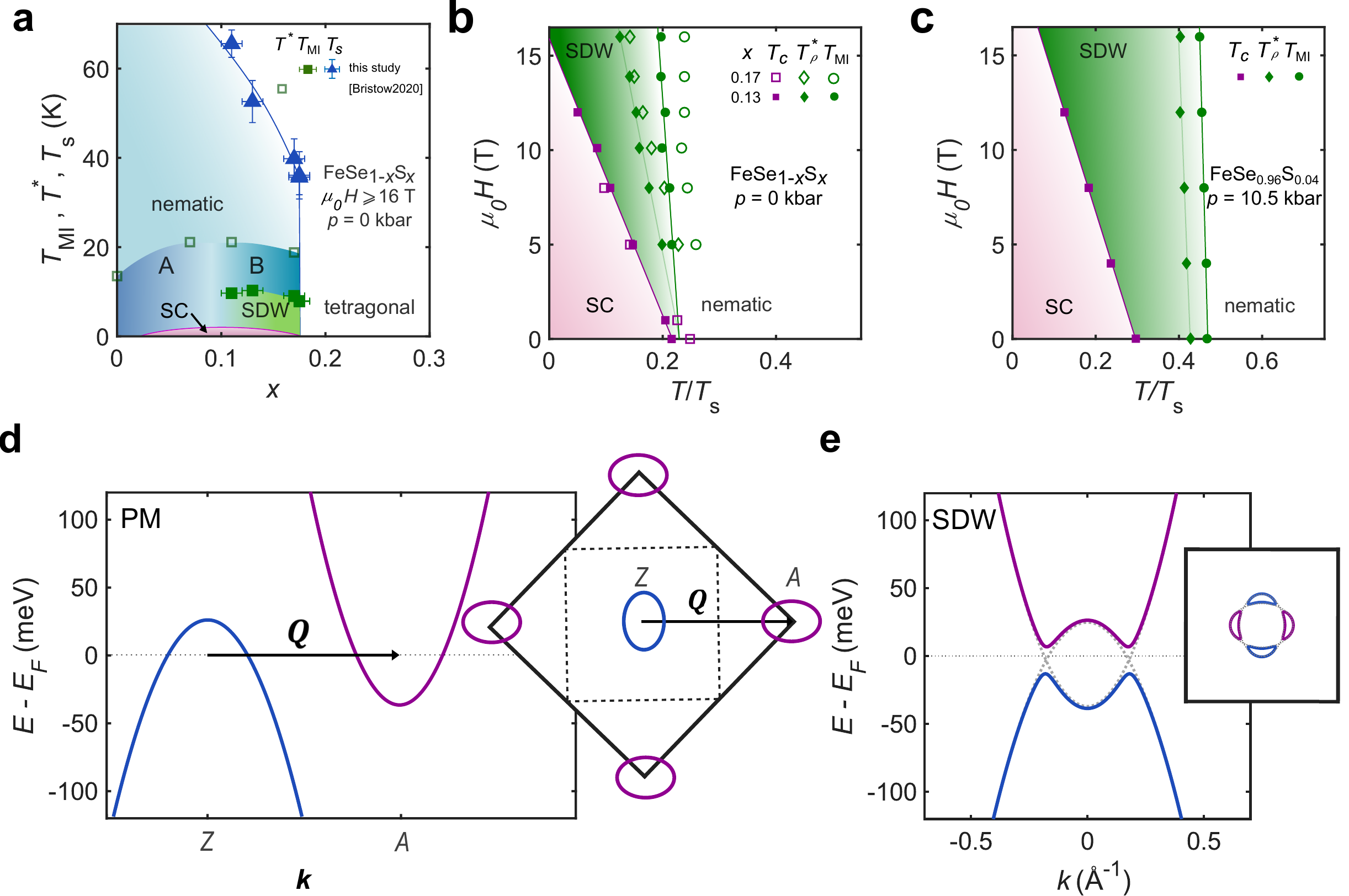}
	\caption{\textbf{Phase diagrams of FeSe$_{1-x}$S$_{x}$.} \textbf{(a)} Phase diagram, $T-x$, as a function of sulfur substitution, $x$, in the presence of magnetic field. Systems investigated in this study, $x=0.11-0.17(5)$ (blue triangles) lie close to the border of the nematic electronic phase indicated by $T_{\rm s}$, which is suppressed close to $x\sim0.18$. The pink region represents residual superconductivity in high magnetic fields. The SDW boundaries are constructed based on transport data in 16~T shown in Supplementary Fig.~S2 and Supplementary Fig.~S3 (solid green squares). Boundary at $T^*$ reflects the change in resistivity slope in 34~T after Ref.~\cite{Bristow2020} (open green squares).    
     \textbf{(b)} Magnetic field-temperature  phase diagram in reduced temperature units, $t=T/T_{\rm s}$, for samples S1 ($x\sim0.13$) and S2 ($x\sim0.17$) at ambient pressure based on transport studies. Solid lines are guides to the eye based on linear fits to the different electronic phase boundaries of sample S1. \textbf{(c)} Equivalent phase diagram to that in (b) but for the case of sample P1 ($x\sim0.04$) under an applied pressure of $p=10.5$~kbar. The extracted temperatures have a systematic error estimated to be $~0.5$~K, due to temperature lag between the sample and the temperature sensors.
       {\textbf{(d)} A schematic
   two-band structure representation showing a hole band (blue) 
      and an electron band (purple).
      The black arrow indicates the nesting vector, ${\bf Q}$, vector. The inset shows the Fermi surface with a hole pocket at the center and
   an electron pocket (in the 2~Fe/unit cell) at the corner of the Brillouin zone (solid line). The dashed lines indicate the magnetic Brillouin zone. 
   \textbf{(e)} The reconstructed band structure and the Fermi surface (inset)
   in the presence of an SDW order.}
   }
   \label{FeSeS13pc_B_T_phase_diagram}

   \end{figure*}

{\bf Comparison with FeSe$_{0.96}$S$_{0.04}$, with similar $T_{\rm s}$, but tuned by applied pressure.}
Figure~\ref{Fig2_FeSeS4pc_pressure} shows the transport and TDO measurements
of two samples of FeSe$_{1-x}$S$_{x}$ with $x\sim0.04$
under applied hydrostatic pressures of $p=10-12$\,kbar.
The resistivity of sample P1 
shows a change in slope
at the nematic transition temperature,  $T_{s}\sim33.6$\,K, which is similar to the samples S1 and S2 measured at ambient pressure. In this case, we observe that the resistivity of P1 has
a well-defined upturn at $T_{\rm MI}$ that defines the
development of the SDW phase, similar to FeSe \cite{Terashima2015}, 
even in the absence of the magnetic field, as shown in Figure~\ref{Fig2_FeSeS4pc_pressure}(a). Sample P1 becomes superconducting below 10\,K and has a high quality with a low temperature residual resistivity,
 $\rho_{\rm xx}({T_{\rm on}} \sim 24 \mu\Omega$cm, and a large $RRR\sim 17$ (see Supplementary Table~S1).
The resistivity behaviour is in good agreement with previous pressure studies
probing the interplane resistivity for a similar composition with $x=0.043$ \cite{Xiang2017}.
With an increasing magnetic field, as superconductivity is suppressed, the characteristic temperature
 $T_{\rm MI}$ shifts slightly, as shown in Supplementary Fig.~S7.

Figure~\ref{Fig2_FeSeS4pc_pressure}(b) shows the temperature dependence of the resonant TDO frequency for sample P2~($x\sim0.04$) under an applied pressure of $p=12(1)$\,kbar. 
Even in the absence of an applied magnetic field, we detect a local peak in the resonant frequency at $T_{p}$, preceding the superconducting transition at $T_{\rm c}$. 
As the applied magnetic field increases, the transition 
at $T_{p}$ hardly changes, 
closely mirroring the  anomaly in resistivity at $T_{\rm MI}$
 (see also the anomalies in the first derivative 
 at $T^{*}_{\rho}$ and $T^{*}_{p}$ under hydrostatic pressure in
 Supplementary Fig.~S7 and Supplementary Fig.~S9, respectively).

Fig.~\ref{Fig2_FeSeS4pc_pressure}(c) 
presents quantum oscillations measured in longitudinal resistivity under pressure for sample P3 ($x=0.04$) measured around 11.5~kbar in high magnetic fields up to 41~T.
We detect a very slow oscillation of frequency close to $F \sim 70$~T      
and an effective mass
$m^*=1.30(2)~m_e$, similar to FeSe \cite{Terashima2015}
 (see Supplementary Fig.~S12).
This suggests a 
Fermi surface reconstruction inside the SDW phase, which leads to a significant reduction in carrier densities and 
the resulting increase in resistivity below $T_{M\rm I}$. Furthermore, such a low frequency is strikingly similar to that observed inside the nematic B phase of FeSe$_{1-x}$S$_x$ (see
Fig.~\ref{Fig1_FeSeS_transport_TDO}(d)),
implying a similar electronic ground state.
Another striking observation under applied pressure is that the magnetoresistance below $T_{\rm MI}$ shows a four-fold increase in 16\,T (see Fig.~\ref{Fig2_FeSeS4pc_pressure}(a)), much larger than that observed at ambient pressure (see Fig.~\ref{Fig1_FeSeS_transport_TDO}(a)).
 The maximum resistivity in 16~T reaches a large value, comparable to that close to 100~K in the zero field (see Supplementary Fig.~S7).
In contrast, within the nematic phase above $T_{\rm MI}$,
the magnetoresistance is significantly reduced for temperatures higher than $T_{\rm MI}$  and is comparable in the two investigated cases [Figure~\ref{Fig2_FeSeS4pc_pressure}(a)) and
	Fig~\ref{Fig1_FeSeS_transport_TDO}(a)].

{\bf DISCUSSION.}

 {\bf Superconducting phase diagrams in a magnetic field.}

To summarize our findings, we have constructed the phase diagram of FeSe$_{1-x}$S$_{x}$ in high fields, as shown in Figure~\ref{FeSeS13pc_B_T_phase_diagram}(a).
Superconductivity is largely suppressed in magnetic fields of 16~T,  and only a residual strip survives across the nematic phase (pink region
in Figure~\ref{FeSeS13pc_B_T_phase_diagram}(a)). 
 Importantly, our study reveals the development of a field-induced incipient SDW phase within the nematic B phase of FeSe$_{1-x}$S$_x$, defined by anomalies detected in transport and TDO measurements (see Figure~\ref{FeSeS13pc_B_T_phase_diagram}(a)).
This incipient SDW phase has close similarities in its electronic manifestations with the SDW stabilized under pressure in FeSe$_{0.96}$S$_{0.04}$, once the nematic phase is suppressed \cite{Zajicek2026}.
To quantify the changes in resistivity between
the two systems, just below $T_{\rm MI}$, 
we consider an exponential dependence of the form $\rho \sim \exp({\Delta_m/k_{\rm B} T})$.
The effective activation energy, $\Delta_{m}$, can be estimated from the slope of the resistivity against the inverse temperature, as shown
in Supplementary Fig.~S13.
Interestingly, we find that $\Delta_{m}$ 
 of FeSe$_{1-x}$S$_{x}$ is almost a factor 10 smaller ($\sim 0.2-0.5$~meV) than that of FeSe$_{0.96}$S$_{0.04}$
and varies roughly with $H^{0.5}$, as detected previously in  underdoped La$_{2-x}$Sr$_x$CuO$_4$ ($x$ = 0.10),
where superconductivity and antiferromagnetism coexist \cite{Lake2002Nature}.
The reduced energy scales establish the incipient 
nature of the SDW inside the nematic B phase of FeSe$_{1-x}$S$_x$ and its sensitivity to magnetic fields.
A schematic representation of
the development of the SDW in a two-band model is shown in Fig.~\ref{FeSeS13pc_B_T_phase_diagram}(d) and (e).

To understand the interplay between the different electronic phases and superconductivity, we compare the $H-T$ phase diagrams  in reduced-temperature units, $t=T/T_{\rm s}$, both at ambient pressure and under applied pressure for different systems.
Figure~\ref{FeSeS13pc_B_T_phase_diagram}(b) shows that the superconducting phase completely obscures the SDW phase 
of FeSe$_{1-x}$S$_{x}$ inside the nematic phase B, and 
magnetic fields above 5~T are essential to reveal its presence (see also Supplementary Fig.~S6).
On the other hand, the stabilization of
the SDW inside the nematic phase is robust in FeSe$_{0.96}$S$_{0.04}$
under applied pressure, with clear signatures of resistivity upturns in the absence
of magnetic field (see Figs.~\ref{FeSeS13pc_B_T_phase_diagram_zero_field} and \ref{FeSeS13pc_B_T_phase_diagram}(c)).
 The suppression rate of superconductivity with an increasing magnetic field, ${dH_{\rm c2}/dT}$, is twice as fast as the shift of the resistivity anomaly associated with the SDW phase (${dH_{\rm m}/dT} \sim 0.22~$K~/T at ambient pressure), as shown in
Fig.~\ref{FeSeS13pc_B_T_phase_diagram}(b).
However, under applied hydrostatic pressure, the SDW phase  occurs at temperatures higher than $T_{\rm c}$, and it is a stable ground state being
hardly affected  by magnetic fields (${dH_{\rm m}/dT} \sim 0.05$~K /T),  as shown in Figure~\ref{FeSeS13pc_B_T_phase_diagram}(c).

The emergence of an incipient SDW phase has consequences on the electronic structure (see Fig.~\ref{FeSeS13pc_B_T_phase_diagram}(e)).
Under applied pressure, quantum oscillations in FeSe change abruptly 
at the onset of the SDW order near the nematic phase border, displaying a small pocket that was associated with a reconstructed Fermi surface \cite{Terashima2015,Terashima2014,Terashima2019}. Our quantum 
oscillation measurements inside the nematic phase reveal a small pocket with a low frequency and small effective mass, tunable by either applied or chemical pressure  (see Supplementary Figs.~S11 and S12).
These comparative studies confirm the presence 
of SDW order inside the nematic B phase of FeSe$_{1-x}$S$_{x}$.
The low frequency pocket
in FeSe$_{1-x}$S$_{x}$
 vanishes with increasing $x$ beyond the nematic endpoint \cite{Coldea2019}.
Furthermore, in high magnetic fields up to 45~T, high frequency oscillations were also observed, but the amplitude of the largest hole pocket, $\delta$, was strongly suppressed inside the nematic phase, which could not be explained by the Lifshitz-Kosevich mass damping term \cite{Coldea2019,Reiss2020}. 
The development of incipient SDW order partially gaps the Fermi surface of FeSe$_{1-x}$S$_{x}$ inside the nematic B phase (see Fig.~\ref{FeSeS13pc_B_T_phase_diagram}(e)),
but the gaps are relatively small (below 1~meV).
 In high magnetic fields, magnetic breakdown via 
tunneling across the SDW gaps could account for the suppressed amplitude of the largest 
Fermi surface pocket, $\delta$ \cite{Coldea2019}. The incipient SDW order, arising from Fermi surface nesting, may 
also be responsible for the anomaly observed in the torque data, as short-range magnetic 
interactions can induce changes in susceptibility anisotropy
\cite{Korshunov2009,Ok2020}. 
Additional magnetic scattering at low temperature could also be responsible for the localization of electrons in thin flakes of FeSe \cite{Farrar2020} and could introduce significant changes in the spectral weight of the electron pockets in the angle-resolved photoemission \cite{Morfoot2025}.

Structurally, both the isovalent substitution and the applied hydrostatic pressure decrease the FeSe lattice parameters (reduction for $10\%$ substitution is equivalent to 3~kbar under applied pressure) \cite{Zajicek2024,Matsuura2017}. Normally, these tuning parameters affect the height of the chalcogen above the conducting Fe planes (see Fig.~\ref{FeSeS13pc_B_T_phase_diagram_zero_field}) in opposite ways, as $z_{p}$ increases under pressure and enhances magnetic interactions, but $z_{x}$ decreases with increasing chemical pressure, $x$ \cite{Moon2010,Yamakaua2017}. 
However, for small values of $x$ or moderate pressures, as in our case, these structural effects suppress
the nematic phase 
in a similar way, thus allowing the stabilization of the SDW-like order (see Fig.~\ref{FeSeS13pc_B_T_phase_diagram_zero_field}). 
Another important difference between the two tuning parameters is the variation in the impurity potential induced by the sulphur for selenium substitution. This isoelectronic substitution could induce a degree of magnetic dilution or frustration, and a reduction in the Stoner enhancement parameter, similar to Ru substitution in iron pnictides \cite{Dhaka2011}.

Resistivity upturns in the absence of a magnetic field have also been observed in doped BaFe$_2$As$_2$, inside the coexistence phase of SDW with superconductivity induced by
various chemical substitutions \cite{Ishida2013}.
As the resistivity upturns at ambient pressure of FeSe$_{1-x}$S$_{x}$ (see Figure \ref{Fig1_FeSeS_transport_TDO}(a)) are not as pronounced  as in the pressure-induced case, and $\Delta_{m}$ is small, these features reflect an incipient form 
of magnetic ordering that is sensitive to applied magnetic fields and can potentially 
coexist with superconductivity.
The electronic reconstruction due to an incipient 
SDW could still preserve a sufficient portion of the Fermi surface for superconducting pairing until the superconducting state is suppressed by the applied magnetic field (see Fig.~\ref{FeSeS13pc_B_T_phase_diagram}(e)).
In such conditions, superconductivity,
with the sign-changing pairing symmetry $s^{\pm}$ promoted by spin fluctuations \cite{Wiecki2017,Wiecki2018}, could coexist with SDW phase over a wider range of parameters
\cite{Cvetkovic2009,Vorontsov2010,Fernandes2010optical}. 
A phase coexistence between the incipient SDW order and superconductivity could influence proposals for the existence of a Bogoliubov Fermi surface for tetragonal FeSe$_{1-x}$S$_{x}$. In such scenarios, ferromagnetic interactions \cite{Peter2023} or toroidal magnetic order \cite{Wu2024} are required, which will lead to a finite residual density of states and anomalous heat capacity \cite{Mizukami2023}.
Additionally, the presence of an incipient SDW 
in the vicinity of the nematic end point could also influence the dominant nematic fluctuations \cite{Islam2024,Nag2025}.

A field-induced incipient SDW has been detected in other unconventional superconductors. In cuprates, magnetic field can induce static antiferromagnetic correlations in the vortex state \cite{Lake2002Nature},
whereas resistivity upturns are
detected in high magnetic fields
\cite{BourgeoisHope2019}.
Recently, pressure-induced superconductivity
in La$_3$Ni$_2$O$_{6.85}$ was found inside the 
orthorhombic structure rather than tetragonal structure \cite{Shi2025}.
In this work, we find that Fermi surface reconstruction
and stabilization of the SDW phase are present inside the nematic phase of iron chalcogenides FeSe$_{1-x}$S$_{x}$ via both applied and chemical pressure. 
Thus, the incipient SDW order acts both as a competing instability that reconstructs the low-energy electronic structure and as a source of collective fluctuations that can contribute to the pairing interaction in unconventional superconductors.

\newpage 
\vspace{0.1cm}

{\bf METHODS}
{\bf Single crystal details.}
Single crystals were grown using the chemical vapour transport method \cite{Bohmer2016, Chareev2013}. 
The estimation of S content, $x$, was 
determined using the structural transition temperature, $T_{\rm s}$, from transport measurements (see Supplementary Figures S1 and S2) and previous studies \cite{Bristow2020}.

{\bf Resistivity measurements.}
Experiments were conducted up to 16~T in a Quantum Design Physical Property Measurement system in Oxford, up to 68~T in a pulsed-field magnet at Laboratoire National des Champs Magnétiques Intenses in Toulouse, and up to 41~T in the resistive magnet at NHMFL in Tallahassee. 
The contacts were soldered to samples with indium
and {\it ac} measurements use a current amplitude of 1\,mA. 
For magnetotransport measurements, all single crystals
were mounted in Hall bar configurations, and both field polarities were used to isolate $\rho_{xx}$ and $\rho_{xy}$ (Supplementary Fig.~S4). 
The residual resistivity ratio was defined as $RRR=\rho_{xx}{(292~\rm K)}/\rho_{xx}(T_{\rm on})$, where the superconducting onset temperature is at $T_{\rm on}$ 
(see Supplementary Table~S1), in good agreement with previous results \cite{Bristow2020}. Resistivity measurements
in pulsed field for sample S2, were corrected
by a constant value of $32.7\mu\Omega$cm to account for
an instrumental offset (see Fig.~\ref{Fig1_FeSeS_transport_TDO}(d)).

{\bf TDO measurements.}
The experimental TDO setup consists of a self-resonating $LC$ oscillator. The AC oscillations in the circuit are sustained by the negative differential resistance of an appropriately biased tunnel diode with the resonant frequency given by $f_0=1/(2\pi \sqrt{L C})$. The sample is then placed in a coil which acts as the inductor in the circuit \cite{van1975tunnel}. 
The inductance, in turn, depends on the skin depth for a normal conductor and the penetration depth inside the superconducting state \cite{VANNETTE2008354}.

{\bf Pressure measurements.}
Pressure measurements were made using 
commercial BeCu pressure cells 
and Daphne 7373 oil as the pressurizing medium \cite{Yamakaua2017}. The in situ pressure was determined mainly using the superconducting transition of tin. The magnetic field is aligned along the $c$-axis for all transport and TDO measurements.

{\bf Torque measurements.}
Magnetic torque measurements 
were performed using piezocantilevers
rotated in constant magnetic fields within the crystallographic $(ac)$-plane.
 For the torque experiments, we employ a current-driven Wheatstone bridge that probes the imbalance between the resistance of two piezocantilevers: one containing the sample and the other without a sample
 to cancel out any potential magnetoresistance effects. In order to calibrate in absolute units, we measure the torque generated by the weight of the sample.

{\bf Extraction of different transition temperatures.}
The nematic transition at $T_{\rm s}$ is defined by a local peak in the second derivative of resistivity against temperature (see Supplementary Fig.~S1 and ref.~\cite{Fisher2018}). 
The uncertainty is estimated from the full width at half maximum (FWHM) of the peak relative to the high-temperature background. The superconducting transition, $T_{\rm c}$, is extracted from the sharp maximum peak in first-order derivative, while the superconducting onset $T_{\rm on}$ and offset $T_{\rm off}$ are determined through the intercepts of linear fits to the data on either side of the transition (see Supplementary Fig.~S5(b,e) and Fig.~S4(c)). $T_{\rm MI}$ denotes the crossover from metallic-like to insulating-like behavior in resistivity due to the development of the SDW phase, and is precisely identified from a local maximum in the second derivative (see Supplementary Fig.~S5(c,f) and Fig.~S7(c)). Analogous features that describe the onset of SDW in TDO frequency response ($T_{\rm p}$ in Fig.~\ref{Fig1_FeSeS_transport_TDO}(e,
Fig.~\ref{Fig2_FeSeS4pc_pressure}(b))
 and magnetic torque ($T_{\rm \tau}$ in Fig.~\ref{Fig1_FeSeS_transport_TDO}(f)) are obtained from local maxima of the raw data. Additionally, $T^{*}_{\rho}$ (or $T^{*}_{p}$) represents the emergent local minimum (or maximum) in the first derivative of resistivity (or frequency) accompanying the field-induced development of incipient SDW (see Supplementary Figs.~S5, S9(c,d) and ref.~\cite{Fisher2018}).  
The extracted temperatures are affected by systematic errors due to differences between 
the sample and thermometer of $0.5$~K.
Additional small errors of less than $0.1$~K may occur from interpolation.

\vspace{0.1 cm}

{\bf ACKNOWLEDGMENTS}
We thank Jorg Schmalian for enlightening discussions.
This work was mainly supported by Engineering and Physical Sciences Research Council (EPSRC)
(EP/I004475/1) and Oxford Centre for Applied Superconductivity and
 the ISABEL project of the European Union’s Horizon's 2020 Research and Innovation Programme Grant Agreement Number No 871106. We also acknowledge the financial support of the John
Fell Fund of the University of Oxford. 
Part of this work was supported by HFML-RU and LNCMI-CNRS, members of the European Magnetic Field Laboratory (EMFL) and by EPSRC (UK) via its membership to the EMFL (grant no. EP/N01085X/1).
We acknowledge the financial support of Oxford University John Fell Fund.
Z.Z. acknowledges financial support from the EPSRC studentship (EP/N509711/1 and EP/R513295/1).
J.S.P  acknowledges financial support from the EPSRC studentship EP/W524311/1, scholarship funding from the Department of Physics, via OxPEG, and the Leathersellers' Scholarship from St. Catherine's College, Oxford.
R.M.A. acknowledges funding from the Margaret Thatcher Scholarship Trust from Somerville College, Oxford.
I.P. acknowledges funding for an iCASE Studentship (EP/W524311/1) and additional sponsorship
from Oxford Instruments. A.A.H. acknowledges support of the Deutsche Forschungsgemeinschaft (DFG; German Research Foundation) under CRC/TRR 288 (Project No. B03).
A.I.C. acknowledges an EPSRC Career Acceleration Fellowship (EP/I004475/1).

{\bf Author Contributions}
IP, ABM, WK, AIC performed transport experiments in Oxford and Toulouse.
RMA, JSP, OS, AIC performed TDO experiments in Oxford and Toulouse.
IP, JSP, AIC performed torque experiments and calibration in Oxford.
ZZ, OS, AIC performed transport and TDO experiments under pressure in Oxford.
AAH synthesized the single crystals for these experiments.
IP, RMA, JSP, ABM, ZZ, OS, WK, AIC performed the data analysis.
IP, RMA, JSP, AIC wrote the paper with contributions and comments from all the authors.
AIC designed the research and supervised research group.

{\bf Corresponding authors.}
To whom correspondence should be addressed. e-mail: ioana.paulescu@physics.ox.ac.uk and
email: amalia.coldea@physics.ox.ac.uk.

{\bf DATA AVAILABILITY.}
All data are available in the manuscript or the supplementary materials. Data supporting the findings of this study will be available through the open-access data archive at the University of Oxford (ORA).

The authors declare no conflict of interest. This article contains supporting information online.

\vspace{0.5cm}

{\bf REFERENCES.} 

\bibliography{biblio}

\begin{thebibliography}{67}%
\makeatletter
\providecommand \@ifxundefined [1]{%
 \@ifx{#1\undefined}
}%
\providecommand \@ifnum [1]{%
 \ifnum #1\expandafter \@firstoftwo
 \else \expandafter \@secondoftwo
 \fi
}%
\providecommand \@ifx [1]{%
 \ifx #1\expandafter \@firstoftwo
 \else \expandafter \@secondoftwo
 \fi
}%
\providecommand \natexlab [1]{#1}%
\providecommand \enquote  [1]{``#1''}%
\providecommand \bibnamefont  [1]{#1}%
\providecommand \bibfnamefont [1]{#1}%
\providecommand \citenamefont [1]{#1}%
\providecommand \href@noop [0]{\@secondoftwo}%
\providecommand \href [0]{\begingroup \@sanitize@url \@href}%
\providecommand \@href[1]{\@@startlink{#1}\@@href}%
\providecommand \@@href[1]{\endgroup#1\@@endlink}%
\providecommand \@sanitize@url [0]{\catcode `\\12\catcode `\$12\catcode
  `\&12\catcode `\#12\catcode `\^12\catcode `\_12\catcode `\%12\relax}%
\providecommand \@@startlink[1]{}%
\providecommand \@@endlink[0]{}%
\providecommand \url  [0]{\begingroup\@sanitize@url \@url }%
\providecommand \@url [1]{\endgroup\@href {#1}{\urlprefix }}%
\providecommand \urlprefix  [0]{URL }%
\providecommand \Eprint [0]{\href }%
\providecommand \doibase [0]{https://doi.org/}%
\providecommand \selectlanguage [0]{\@gobble}%
\providecommand \bibinfo  [0]{\@secondoftwo}%
\providecommand \bibfield  [0]{\@secondoftwo}%
\providecommand \translation [1]{[#1]}%
\providecommand \BibitemOpen [0]{}%
\providecommand \bibitemStop [0]{}%
\providecommand \bibitemNoStop [0]{.\EOS\space}%
\providecommand \EOS [0]{\spacefactor3000\relax}%
\providecommand \BibitemShut  [1]{\csname bibitem#1\endcsname}%
\let\auto@bib@innerbib\@empty
%</preamble>
\bibitem [{\citenamefont {Keimer}\ \emph {et~al.}(2015)\citenamefont {Keimer},
  \citenamefont {Kivelson}, \citenamefont {Norman}, \citenamefont {Uchida},\
  and\ \citenamefont {Zaanen}}]{Keimer2015}%
  \BibitemOpen
  \bibfield  {author} {\bibinfo {author} {\bibfnamefont {B.}~\bibnamefont
  {Keimer}}, \bibinfo {author} {\bibfnamefont {S.~A.}\ \bibnamefont
  {Kivelson}}, \bibinfo {author} {\bibfnamefont {M.~R.}\ \bibnamefont
  {Norman}}, \bibinfo {author} {\bibfnamefont {S.}~\bibnamefont {Uchida}},\
  and\ \bibinfo {author} {\bibfnamefont {J.}~\bibnamefont {Zaanen}},\
  }\bibfield  {title} {\bibinfo {title} {From quantum matter to
  high-temperature superconductivity in copper oxides},\ }\href
  {https://doi.org/10.1038/nature14165} {\bibfield  {journal} {\bibinfo
  {journal} {Nature}\ }\textbf {\bibinfo {volume} {518}},\ \bibinfo {pages}
  {179} (\bibinfo {year} {2015})}\BibitemShut {NoStop}%
\bibitem [{\citenamefont {Shi}\ \emph {et~al.}(2025)\citenamefont {Shi},
  \citenamefont {Peng}, \citenamefont {Li}, \citenamefont {Yang}, \citenamefont
  {Xing}, \citenamefont {Wang}, \citenamefont {Fan}, \citenamefont {Li},
  \citenamefont {Wu}, \citenamefont {Ge}, \citenamefont {Zeng}, \citenamefont
  {Zeng}, \citenamefont {Ying}, \citenamefont {Wu},\ and\ \citenamefont
  {Chen}}]{Shi2025}%
  \BibitemOpen
  \bibfield  {author} {\bibinfo {author} {\bibfnamefont {M.}~\bibnamefont
  {Shi}}, \bibinfo {author} {\bibfnamefont {D.}~\bibnamefont {Peng}}, \bibinfo
  {author} {\bibfnamefont {Y.}~\bibnamefont {Li}}, \bibinfo {author}
  {\bibfnamefont {S.}~\bibnamefont {Yang}}, \bibinfo {author} {\bibfnamefont
  {Z.}~\bibnamefont {Xing}}, \bibinfo {author} {\bibfnamefont {Y.}~\bibnamefont
  {Wang}}, \bibinfo {author} {\bibfnamefont {K.}~\bibnamefont {Fan}}, \bibinfo
  {author} {\bibfnamefont {H.}~\bibnamefont {Li}}, \bibinfo {author}
  {\bibfnamefont {R.}~\bibnamefont {Wu}}, \bibinfo {author} {\bibfnamefont
  {B.}~\bibnamefont {Ge}}, \bibinfo {author} {\bibfnamefont {Z.}~\bibnamefont
  {Zeng}}, \bibinfo {author} {\bibfnamefont {Q.}~\bibnamefont {Zeng}}, \bibinfo
  {author} {\bibfnamefont {J.}~\bibnamefont {Ying}}, \bibinfo {author}
  {\bibfnamefont {T.}~\bibnamefont {Wu}},\ and\ \bibinfo {author}
  {\bibfnamefont {X.}~\bibnamefont {Chen}},\ }\bibfield  {title} {\bibinfo
  {title} {{Spin density wave rather than tetragonal structure is prerequisite
  for superconductivity in La3Ni2O7-$\delta$}},\ }\href
  {https://doi.org/10.1038/s41467-025-63701-x} {\bibfield  {journal} {\bibinfo
  {journal} {Nature Communications}\ }\textbf {\bibinfo {volume} {16}},\
  \bibinfo {pages} {9141} (\bibinfo {year} {2025})}\BibitemShut {NoStop}%
\bibitem [{\citenamefont {Maple}(1995)}]{Maple1995}%
  \BibitemOpen
  \bibfield  {author} {\bibinfo {author} {\bibfnamefont {M.}~\bibnamefont
  {Maple}},\ }\bibfield  {title} {\bibinfo {title} {Interplay between
  superconductivity and magnetism},\ }\href
  {https://doi.org/https://doi.org/10.1016/0921-4526(95)00031-4} {\bibfield
  {journal} {\bibinfo  {journal} {Physica B: Condensed Matter}\ }\textbf
  {\bibinfo {volume} {215}},\ \bibinfo {pages} {110} (\bibinfo {year}
  {1995})}\BibitemShut {NoStop}%
\bibitem [{\citenamefont {Fernandes}\ \emph {et~al.}(2022)\citenamefont
  {Fernandes}, \citenamefont {Coldea}, \citenamefont {Ding}, \citenamefont
  {Fisher}, \citenamefont {Hirschfeld},\ and\ \citenamefont
  {Kotliar}}]{Fernandes2022}%
  \BibitemOpen
  \bibfield  {author} {\bibinfo {author} {\bibfnamefont {R.~M.}\ \bibnamefont
  {Fernandes}}, \bibinfo {author} {\bibfnamefont {A.~I.}\ \bibnamefont
  {Coldea}}, \bibinfo {author} {\bibfnamefont {H.}~\bibnamefont {Ding}},
  \bibinfo {author} {\bibfnamefont {I.~R.}\ \bibnamefont {Fisher}}, \bibinfo
  {author} {\bibfnamefont {P.~J.}\ \bibnamefont {Hirschfeld}},\ and\ \bibinfo
  {author} {\bibfnamefont {G.}~\bibnamefont {Kotliar}},\ }\bibfield  {title}
  {\bibinfo {title} {{Iron pnictides and chalcogenides: a new paradigm for
  superconductivity}},\ }\href {https://doi.org/10.1038/s41586-021-04073-2}
  {\bibfield  {journal} {\bibinfo  {journal} {Nature}\ }\textbf {\bibinfo
  {volume} {601}},\ \bibinfo {pages} {35} (\bibinfo {year} {2022})}\BibitemShut
  {NoStop}%
\bibitem [{\citenamefont {{Sprau}}\ \emph {et~al.}(2016)\citenamefont
  {{Sprau}}, \citenamefont {{Kostin}}, \citenamefont {{Kreisel}}, \citenamefont
  {{B{\"o}hmer}}, \citenamefont {{Taufour}}, \citenamefont {{Canfield}},
  \citenamefont {{Mukherjee}}, \citenamefont {{Hirschfeld}}, \citenamefont
  {{Andersen}},\ and\ \citenamefont {{S{\'e}amus Davis}}}]{Sprau2016}%
  \BibitemOpen
  \bibfield  {author} {\bibinfo {author} {\bibfnamefont {P.~O.}\ \bibnamefont
  {{Sprau}}}, \bibinfo {author} {\bibfnamefont {A.}~\bibnamefont {{Kostin}}},
  \bibinfo {author} {\bibfnamefont {A.}~\bibnamefont {{Kreisel}}}, \bibinfo
  {author} {\bibfnamefont {A.~E.}\ \bibnamefont {{B{\"o}hmer}}}, \bibinfo
  {author} {\bibfnamefont {V.}~\bibnamefont {{Taufour}}}, \bibinfo {author}
  {\bibfnamefont {P.~C.}\ \bibnamefont {{Canfield}}}, \bibinfo {author}
  {\bibfnamefont {S.}~\bibnamefont {{Mukherjee}}}, \bibinfo {author}
  {\bibfnamefont {P.~J.}\ \bibnamefont {{Hirschfeld}}}, \bibinfo {author}
  {\bibfnamefont {B.~M.}\ \bibnamefont {{Andersen}}},\ and\ \bibinfo {author}
  {\bibfnamefont {J.~C.}\ \bibnamefont {{S{\'e}amus Davis}}},\ }\bibfield
  {title} {\bibinfo {title} {{Discovery of Orbital-Selective Cooper Pairing in
  FeSe}},\ }\href {https://arxiv.org/abs/1611.02134} {\bibfield  {journal}
  {\bibinfo  {journal} {Science}\ }\textbf {\bibinfo {volume} {357}},\ \bibinfo
  {pages} {75} (\bibinfo {year} {2016})}\BibitemShut {NoStop}%
\bibitem [{\citenamefont {Nag}\ \emph {et~al.}(2025)\citenamefont {Nag},
  \citenamefont {Scott}, \citenamefont {de~Carvalho}, \citenamefont {Byland},
  \citenamefont {Yang}, \citenamefont {Walker}, \citenamefont {Greenberg},
  \citenamefont {Klavins}, \citenamefont {Miranda}, \citenamefont {Gozar},
  \citenamefont {Taufour}, \citenamefont {Fernandes},\ and\ \citenamefont {{da
  Silva Neto}}}]{Nag2025}%
  \BibitemOpen
  \bibfield  {author} {\bibinfo {author} {\bibfnamefont {P.~K.}\ \bibnamefont
  {Nag}}, \bibinfo {author} {\bibfnamefont {K.}~\bibnamefont {Scott}}, \bibinfo
  {author} {\bibfnamefont {V.~S.}\ \bibnamefont {de~Carvalho}}, \bibinfo
  {author} {\bibfnamefont {J.~K.}\ \bibnamefont {Byland}}, \bibinfo {author}
  {\bibfnamefont {X.}~\bibnamefont {Yang}}, \bibinfo {author} {\bibfnamefont
  {M.}~\bibnamefont {Walker}}, \bibinfo {author} {\bibfnamefont {A.~G.}\
  \bibnamefont {Greenberg}}, \bibinfo {author} {\bibfnamefont {P.}~\bibnamefont
  {Klavins}}, \bibinfo {author} {\bibfnamefont {E.}~\bibnamefont {Miranda}},
  \bibinfo {author} {\bibfnamefont {A.}~\bibnamefont {Gozar}}, \bibinfo
  {author} {\bibfnamefont {V.}~\bibnamefont {Taufour}}, \bibinfo {author}
  {\bibfnamefont {R.~M.}\ \bibnamefont {Fernandes}},\ and\ \bibinfo {author}
  {\bibfnamefont {E.~H.}\ \bibnamefont {{da Silva Neto}}},\ }\bibfield  {title}
  {\bibinfo {title} {{Highly anisotropic superconducting gap near the nematic
  quantum critical point of FeSe$_{1-x}$S$_x$}},\ }\href
  {https://doi.org/10.1038/s41567-024-02683-x} {\bibfield  {journal} {\bibinfo
  {journal} {Nature Physics}\ }\textbf {\bibinfo {volume} {21}},\ \bibinfo
  {pages} {89} (\bibinfo {year} {2025})}\BibitemShut {NoStop}%
\bibitem [{\citenamefont {Dai}\ \emph {et~al.}(2012)\citenamefont {Dai},
  \citenamefont {Hu},\ and\ \citenamefont {Dagotto}}]{Dai2012}%
  \BibitemOpen
  \bibfield  {author} {\bibinfo {author} {\bibfnamefont {P.}~\bibnamefont
  {Dai}}, \bibinfo {author} {\bibfnamefont {J.}~\bibnamefont {Hu}},\ and\
  \bibinfo {author} {\bibfnamefont {E.}~\bibnamefont {Dagotto}},\ }\bibfield
  {title} {\bibinfo {title} {{Magnetism and its microscopic origin in
  iron-based high-temperature superconductors}},\ }\href
  {https://doi.org/10.1038/nphys2438} {\bibfield  {journal} {\bibinfo
  {journal} {Nature Physics}\ }\textbf {\bibinfo {volume} {8}},\ \bibinfo
  {pages} {709} (\bibinfo {year} {2012})}\BibitemShut {NoStop}%
\bibitem [{\citenamefont {{de Medici}}\ \emph {et~al.}(2014)\citenamefont {{de
  Medici}}, \citenamefont {Giovannetti},\ and\ \citenamefont
  {Capone}}]{deMedici2014}%
  \BibitemOpen
  \bibfield  {author} {\bibinfo {author} {\bibfnamefont {L.}~\bibnamefont {{de
  Medici}}}, \bibinfo {author} {\bibfnamefont {G.}~\bibnamefont
  {Giovannetti}},\ and\ \bibinfo {author} {\bibfnamefont {M.}~\bibnamefont
  {Capone}},\ }\bibfield  {title} {\bibinfo {title} {{Selective Mott Physics as
  a Key to Iron Superconductors}},\ }\href
  {https://doi.org/10.1103/PhysRevLett.112.177001} {\bibfield  {journal}
  {\bibinfo  {journal} {Phys. Rev. Lett.}\ }\textbf {\bibinfo {volume} {112}},\
  \bibinfo {pages} {177001} (\bibinfo {year} {2014})}\BibitemShut {NoStop}%
\bibitem [{\citenamefont {Dai}(2015)}]{Dai2015}%
  \BibitemOpen
  \bibfield  {author} {\bibinfo {author} {\bibfnamefont {P.}~\bibnamefont
  {Dai}},\ }\bibfield  {title} {\bibinfo {title} {Antiferromagnetic order and
  spin dynamics in iron-based superconductors},\ }\href
  {https://doi.org/10.1103/RevModPhys.87.855} {\bibfield  {journal} {\bibinfo
  {journal} {Rev. Mod. Phys.}\ }\textbf {\bibinfo {volume} {87}},\ \bibinfo
  {pages} {855} (\bibinfo {year} {2015})}\BibitemShut {NoStop}%
\bibitem [{\citenamefont {Mazin}\ \emph {et~al.}(2008)\citenamefont {Mazin},
  \citenamefont {Singh}, \citenamefont {Johannes},\ and\ \citenamefont
  {Du}}]{Mazin2008}%
  \BibitemOpen
  \bibfield  {author} {\bibinfo {author} {\bibfnamefont {I.~I.}\ \bibnamefont
  {Mazin}}, \bibinfo {author} {\bibfnamefont {D.~J.}\ \bibnamefont {Singh}},
  \bibinfo {author} {\bibfnamefont {M.~D.}\ \bibnamefont {Johannes}},\ and\
  \bibinfo {author} {\bibfnamefont {M.~H.}\ \bibnamefont {Du}},\ }\bibfield
  {title} {\bibinfo {title} {{Unconventional Superconductivity with a Sign
  Reversal in the Order Parameter of
  ${\mathrm{LaFeAsO}}_{1\ensuremath{-}x}{\mathrm{F}}_{x}$}},\ }\href
  {https://doi.org/10.1103/PhysRevLett.101.057003} {\bibfield  {journal}
  {\bibinfo  {journal} {Phys. Rev. Lett.}\ }\textbf {\bibinfo {volume} {101}},\
  \bibinfo {pages} {057003} (\bibinfo {year} {2008})}\BibitemShut {NoStop}%
\bibitem [{\citenamefont {Terashima}\ \emph {et~al.}(2009)\citenamefont
  {Terashima}, \citenamefont {Sekiba}, \citenamefont {Bowen}, \citenamefont
  {Nakayama}, \citenamefont {Kawahara}, \citenamefont {Sato}, \citenamefont
  {Richard}, \citenamefont {Xu}, \citenamefont {Li}, \citenamefont {Cao},
  \citenamefont {Xu}, \citenamefont {Ding},\ and\ \citenamefont
  {Takahashi}}]{Terashima2009}%
  \BibitemOpen
  \bibfield  {author} {\bibinfo {author} {\bibfnamefont {K.}~\bibnamefont
  {Terashima}}, \bibinfo {author} {\bibfnamefont {Y.}~\bibnamefont {Sekiba}},
  \bibinfo {author} {\bibfnamefont {J.~H.}\ \bibnamefont {Bowen}}, \bibinfo
  {author} {\bibfnamefont {K.}~\bibnamefont {Nakayama}}, \bibinfo {author}
  {\bibfnamefont {T.}~\bibnamefont {Kawahara}}, \bibinfo {author}
  {\bibfnamefont {T.}~\bibnamefont {Sato}}, \bibinfo {author} {\bibfnamefont
  {P.}~\bibnamefont {Richard}}, \bibinfo {author} {\bibfnamefont {Y.-M.}\
  \bibnamefont {Xu}}, \bibinfo {author} {\bibfnamefont {L.~J.}\ \bibnamefont
  {Li}}, \bibinfo {author} {\bibfnamefont {G.~H.}\ \bibnamefont {Cao}},
  \bibinfo {author} {\bibfnamefont {Z.-A.}\ \bibnamefont {Xu}}, \bibinfo
  {author} {\bibfnamefont {H.}~\bibnamefont {Ding}},\ and\ \bibinfo {author}
  {\bibfnamefont {T.}~\bibnamefont {Takahashi}},\ }\bibfield  {title} {\bibinfo
  {title} {Fermi surface nesting induced strong pairing in iron-based
  superconductors},\ }\href {https://doi.org/10.1073/pnas.0900469106}
  {\bibfield  {journal} {\bibinfo  {journal} {Proceedings of the National
  Academy of Sciences}\ }\textbf {\bibinfo {volume} {106}},\ \bibinfo {pages}
  {7330} (\bibinfo {year} {2009})}\BibitemShut {NoStop}%
\bibitem [{\citenamefont {Chu}\ \emph {et~al.}(2009)\citenamefont {Chu},
  \citenamefont {Analytis}, \citenamefont {Kucharczyk},\ and\ \citenamefont
  {Fisher}}]{Chu2009}%
  \BibitemOpen
  \bibfield  {author} {\bibinfo {author} {\bibfnamefont {J.-H.}\ \bibnamefont
  {Chu}}, \bibinfo {author} {\bibfnamefont {J.~G.}\ \bibnamefont {Analytis}},
  \bibinfo {author} {\bibfnamefont {C.}~\bibnamefont {Kucharczyk}},\ and\
  \bibinfo {author} {\bibfnamefont {I.~R.}\ \bibnamefont {Fisher}},\ }\bibfield
   {title} {\bibinfo {title} {{Determination of the phase diagram of the
  electron-doped superconductor
  $\text{Ba}{({\text{Fe}}_{1\ensuremath{-}x}{\text{Co}}_{x})}_{2}{\text{As}}_{2}$}},\
  }\href {https://doi.org/10.1103/PhysRevB.79.014506} {\bibfield  {journal}
  {\bibinfo  {journal} {Phys. Rev. B}\ }\textbf {\bibinfo {volume} {79}},\
  \bibinfo {pages} {014506} (\bibinfo {year} {2009})}\BibitemShut {NoStop}%
\bibitem [{\citenamefont {Kimber}\ \emph {et~al.}(2009)\citenamefont {Kimber},
  \citenamefont {Kreyssig}, \citenamefont {Zhang}, \citenamefont {Jeschke},
  \citenamefont {Valent{\'i}}, \citenamefont {Yokaichiya}, \citenamefont
  {Colombier}, \citenamefont {Yan}, \citenamefont {Hansen}, \citenamefont
  {Chatterji}, \citenamefont {McQueeney}, \citenamefont {Canfield},
  \citenamefont {Goldman},\ and\ \citenamefont {Argyriou}}]{Kimber2009}%
  \BibitemOpen
  \bibfield  {author} {\bibinfo {author} {\bibfnamefont {S.~A.~J.}\
  \bibnamefont {Kimber}}, \bibinfo {author} {\bibfnamefont {A.}~\bibnamefont
  {Kreyssig}}, \bibinfo {author} {\bibfnamefont {Y.-Z.}\ \bibnamefont {Zhang}},
  \bibinfo {author} {\bibfnamefont {H.~O.}\ \bibnamefont {Jeschke}}, \bibinfo
  {author} {\bibfnamefont {R.}~\bibnamefont {Valent{\'i}}}, \bibinfo {author}
  {\bibfnamefont {F.}~\bibnamefont {Yokaichiya}}, \bibinfo {author}
  {\bibfnamefont {E.}~\bibnamefont {Colombier}}, \bibinfo {author}
  {\bibfnamefont {J.}~\bibnamefont {Yan}}, \bibinfo {author} {\bibfnamefont
  {T.~C.}\ \bibnamefont {Hansen}}, \bibinfo {author} {\bibfnamefont
  {T.}~\bibnamefont {Chatterji}}, \bibinfo {author} {\bibfnamefont {R.~J.}\
  \bibnamefont {McQueeney}}, \bibinfo {author} {\bibfnamefont {P.~C.}\
  \bibnamefont {Canfield}}, \bibinfo {author} {\bibfnamefont {A.~I.}\
  \bibnamefont {Goldman}},\ and\ \bibinfo {author} {\bibfnamefont {D.~N.}\
  \bibnamefont {Argyriou}},\ }\bibfield  {title} {\bibinfo {title}
  {{Similarities between structural distortions under pressure and chemical
  doping in superconducting BaFe$_2$As$_2$}},\ }\href
  {https://doi.org/10.1038/nmat2443} {\bibfield  {journal} {\bibinfo  {journal}
  {Nature Materials}\ }\textbf {\bibinfo {volume} {8}},\ \bibinfo {pages} {471}
  (\bibinfo {year} {2009})}\BibitemShut {NoStop}%
\bibitem [{\citenamefont {Cvetkovic}\ and\ \citenamefont
  {Tesanovic}(2009)}]{Cvetkovic2009}%
  \BibitemOpen
  \bibfield  {author} {\bibinfo {author} {\bibfnamefont {V.}~\bibnamefont
  {Cvetkovic}}\ and\ \bibinfo {author} {\bibfnamefont {Z.}~\bibnamefont
  {Tesanovic}},\ }\bibfield  {title} {\bibinfo {title} {Valley density-wave and
  multiband superconductivity in iron-based pnictide superconductors},\ }\href
  {https://doi.org/10.1103/PhysRevB.80.024512} {\bibfield  {journal} {\bibinfo
  {journal} {Phys. Rev. B}\ }\textbf {\bibinfo {volume} {80}},\ \bibinfo
  {pages} {024512} (\bibinfo {year} {2009})}\BibitemShut {NoStop}%
\bibitem [{\citenamefont {Fernandes}\ and\ \citenamefont
  {Schmalian}(2010{\natexlab{a}})}]{Fernandes2010}%
  \BibitemOpen
  \bibfield  {author} {\bibinfo {author} {\bibfnamefont {R.~M.}\ \bibnamefont
  {Fernandes}}\ and\ \bibinfo {author} {\bibfnamefont {J.}~\bibnamefont
  {Schmalian}},\ }\bibfield  {title} {\bibinfo {title} {{Competing order and
  nature of the pairing state in the iron pnictides}},\ }\href
  {https://doi.org/10.1103/PhysRevB.82.014521} {\bibfield  {journal} {\bibinfo
  {journal} {Phys. Rev. B}\ }\textbf {\bibinfo {volume} {82}},\ \bibinfo
  {pages} {014521} (\bibinfo {year} {2010}{\natexlab{a}})}\BibitemShut
  {NoStop}%
\bibitem [{\citenamefont {Vorontsov}\ \emph {et~al.}(2010)\citenamefont
  {Vorontsov}, \citenamefont {Vavilov},\ and\ \citenamefont
  {Chubukov}}]{Vorontsov2010}%
  \BibitemOpen
  \bibfield  {author} {\bibinfo {author} {\bibfnamefont {A.~B.}\ \bibnamefont
  {Vorontsov}}, \bibinfo {author} {\bibfnamefont {M.~G.}\ \bibnamefont
  {Vavilov}},\ and\ \bibinfo {author} {\bibfnamefont {A.~V.}\ \bibnamefont
  {Chubukov}},\ }\bibfield  {title} {\bibinfo {title} {Superconductivity and
  spin-density waves in multiband metals},\ }\href
  {https://doi.org/10.1103/PhysRevB.81.174538} {\bibfield  {journal} {\bibinfo
  {journal} {Phys. Rev. B}\ }\textbf {\bibinfo {volume} {81}},\ \bibinfo
  {pages} {174538} (\bibinfo {year} {2010})}\BibitemShut {NoStop}%
\bibitem [{\citenamefont {Hirschfeld}\ \emph {et~al.}(2011)\citenamefont
  {Hirschfeld}, \citenamefont {Korshunov},\ and\ \citenamefont
  {Mazin}}]{Hirschfeld2011}%
  \BibitemOpen
  \bibfield  {author} {\bibinfo {author} {\bibfnamefont {P.~J.}\ \bibnamefont
  {Hirschfeld}}, \bibinfo {author} {\bibfnamefont {M.~M.}\ \bibnamefont
  {Korshunov}},\ and\ \bibinfo {author} {\bibfnamefont {I.~I.}\ \bibnamefont
  {Mazin}},\ }\bibfield  {title} {\bibinfo {title} {{Gap symmetry and structure
  of Fe-based superconductors}},\ }\href
  {https://doi.org/10.1088/0034-4885/74/12/124508} {\bibfield  {journal}
  {\bibinfo  {journal} {Reports on Progress in Physics}\ }\textbf {\bibinfo
  {volume} {74}},\ \bibinfo {pages} {124508} (\bibinfo {year}
  {2011})}\BibitemShut {NoStop}%
\bibitem [{\citenamefont {Glasbrenner}\ \emph {et~al.}(2015)\citenamefont
  {Glasbrenner}, \citenamefont {Mazin}, \citenamefont {Jeschke}, \citenamefont
  {Hirschfeld}, \citenamefont {Fernandes},\ and\ \citenamefont
  {Valentí}}]{Glasbrenner2015}%
  \BibitemOpen
  \bibfield  {author} {\bibinfo {author} {\bibfnamefont {J.~K.}\ \bibnamefont
  {Glasbrenner}}, \bibinfo {author} {\bibfnamefont {I.~I.}\ \bibnamefont
  {Mazin}}, \bibinfo {author} {\bibfnamefont {H.~O.}\ \bibnamefont {Jeschke}},
  \bibinfo {author} {\bibfnamefont {P.~J.}\ \bibnamefont {Hirschfeld}},
  \bibinfo {author} {\bibfnamefont {R.~M.}\ \bibnamefont {Fernandes}},\ and\
  \bibinfo {author} {\bibfnamefont {R.}~\bibnamefont {Valentí}},\ }\bibfield
  {title} {\bibinfo {title} {Effect of magnetic frustration on nematicity and
  superconductivity in iron chalcogenides},\ }\href
  {https://doi.org/10.1038/nphys3434} {\bibfield  {journal} {\bibinfo
  {journal} {Nature Physics}\ }\textbf {\bibinfo {volume} {11}},\ \bibinfo
  {pages} {953} (\bibinfo {year} {2015})}\BibitemShut {NoStop}%
\bibitem [{\citenamefont {Coldea}\ and\ \citenamefont
  {Watson}(2018)}]{Amalia2018}%
  \BibitemOpen
  \bibfield  {author} {\bibinfo {author} {\bibfnamefont {A.~I.}\ \bibnamefont
  {Coldea}}\ and\ \bibinfo {author} {\bibfnamefont {M.~D.}\ \bibnamefont
  {Watson}},\ }\bibfield  {title} {\bibinfo {title} {The key ingredients of the
  electronic structure of {FeSe}},\ }\href
  {https://doi.org/10.1146/annurev-conmatphys-033117-054137} {\bibfield
  {journal} {\bibinfo  {journal} {Annual Review of Condensed Matter Physics}\
  }\textbf {\bibinfo {volume} {9}},\ \bibinfo {pages} {125–146} (\bibinfo
  {year} {2018})}\BibitemShut {NoStop}%
\bibitem [{\citenamefont {Coldea}(2021)}]{Amalia2021}%
  \BibitemOpen
  \bibfield  {author} {\bibinfo {author} {\bibfnamefont {A.~I.}\ \bibnamefont
  {Coldea}},\ }\bibfield  {title} {\bibinfo {title} {Electronic nematic states
  tuned by isoelectronic substitution in bulk
  {FeSe\textsubscript{1-x}S\textsubscript{x}}},\ }\bibfield  {journal}
  {\bibinfo  {journal} {Frontiers in Physics}\ }\textbf {\bibinfo {volume}
  {8}},\ \href {https://doi.org/10.3389/fphy.2020.594500}
  {10.3389/fphy.2020.594500} (\bibinfo {year} {2021})\BibitemShut {NoStop}%
\bibitem [{\citenamefont {B\"ohmer}\ \emph {et~al.}(2016)\citenamefont
  {B\"ohmer}, \citenamefont {Taufour}, \citenamefont {Straszheim},
  \citenamefont {Wolf},\ and\ \citenamefont {Canfield}}]{Bohmer2016}%
  \BibitemOpen
  \bibfield  {author} {\bibinfo {author} {\bibfnamefont {A.~E.}\ \bibnamefont
  {B\"ohmer}}, \bibinfo {author} {\bibfnamefont {V.}~\bibnamefont {Taufour}},
  \bibinfo {author} {\bibfnamefont {W.~E.}\ \bibnamefont {Straszheim}},
  \bibinfo {author} {\bibfnamefont {T.}~\bibnamefont {Wolf}},\ and\ \bibinfo
  {author} {\bibfnamefont {P.~C.}\ \bibnamefont {Canfield}},\ }\bibfield
  {title} {\bibinfo {title} {Variation of transition temperatures and residual
  resistivity ratio in vapor-grown {FeSe}},\ }\href
  {https://doi.org/10.1103/PhysRevB.94.024526} {\bibfield  {journal} {\bibinfo
  {journal} {Phys. Rev. B}\ }\textbf {\bibinfo {volume} {94}},\ \bibinfo
  {pages} {024526} (\bibinfo {year} {2016})}\BibitemShut {NoStop}%
\bibitem [{\citenamefont {Wiecki}\ \emph {et~al.}(2018)\citenamefont {Wiecki},
  \citenamefont {Rana}, \citenamefont {B\"ohmer}, \citenamefont {Lee},
  \citenamefont {Bud'ko}, \citenamefont {Canfield},\ and\ \citenamefont
  {Furukawa}}]{Wiecki2018}%
  \BibitemOpen
  \bibfield  {author} {\bibinfo {author} {\bibfnamefont {P.}~\bibnamefont
  {Wiecki}}, \bibinfo {author} {\bibfnamefont {K.}~\bibnamefont {Rana}},
  \bibinfo {author} {\bibfnamefont {A.~E.}\ \bibnamefont {B\"ohmer}}, \bibinfo
  {author} {\bibfnamefont {Y.}~\bibnamefont {Lee}}, \bibinfo {author}
  {\bibfnamefont {S.~L.}\ \bibnamefont {Bud'ko}}, \bibinfo {author}
  {\bibfnamefont {P.~C.}\ \bibnamefont {Canfield}},\ and\ \bibinfo {author}
  {\bibfnamefont {Y.}~\bibnamefont {Furukawa}},\ }\bibfield  {title} {\bibinfo
  {title} {{Persistent correlation between superconductivity and
  antiferromagnetic fluctuations near a nematic quantum critical point in
  FeSe\textsubscript{1-x}S\textsubscript{x}}},\ }\href
  {https://doi.org/10.1103/PhysRevB.98.020507} {\bibfield  {journal} {\bibinfo
  {journal} {Phys. Rev. B}\ }\textbf {\bibinfo {volume} {98}},\ \bibinfo
  {pages} {020507} (\bibinfo {year} {2018})}\BibitemShut {NoStop}%
\bibitem [{\citenamefont {Wang}\ \emph {et~al.}(2016)\citenamefont {Wang},
  \citenamefont {Shen}, \citenamefont {Pan}, \citenamefont {Zhang},
  \citenamefont {Ikeuchi}, \citenamefont {Iida}, \citenamefont {Christianson},
  \citenamefont {Walker}, \citenamefont {Adroja}, \citenamefont {Abdel-Hafiez},
  \citenamefont {Chen}, \citenamefont {Chareev}, \citenamefont {Vasiliev},\
  and\ \citenamefont {Zhao}}]{Wang2016}%
  \BibitemOpen
  \bibfield  {author} {\bibinfo {author} {\bibfnamefont {Q.}~\bibnamefont
  {Wang}}, \bibinfo {author} {\bibfnamefont {Y.}~\bibnamefont {Shen}}, \bibinfo
  {author} {\bibfnamefont {B.}~\bibnamefont {Pan}}, \bibinfo {author}
  {\bibfnamefont {X.}~\bibnamefont {Zhang}}, \bibinfo {author} {\bibfnamefont
  {K.}~\bibnamefont {Ikeuchi}}, \bibinfo {author} {\bibfnamefont
  {K.}~\bibnamefont {Iida}}, \bibinfo {author} {\bibfnamefont {A.~D.}\
  \bibnamefont {Christianson}}, \bibinfo {author} {\bibfnamefont {H.~C.}\
  \bibnamefont {Walker}}, \bibinfo {author} {\bibfnamefont {D.~T.}\
  \bibnamefont {Adroja}}, \bibinfo {author} {\bibfnamefont {M.}~\bibnamefont
  {Abdel-Hafiez}}, \bibinfo {author} {\bibfnamefont {X.}~\bibnamefont {Chen}},
  \bibinfo {author} {\bibfnamefont {D.~A.}\ \bibnamefont {Chareev}}, \bibinfo
  {author} {\bibfnamefont {A.~N.}\ \bibnamefont {Vasiliev}},\ and\ \bibinfo
  {author} {\bibfnamefont {J.}~\bibnamefont {Zhao}},\ }\bibfield  {title}
  {\bibinfo {title} {{Magnetic ground state of FeSe}},\ }\href
  {https://doi.org/10.1038/ncomms12182} {\bibfield  {journal} {\bibinfo
  {journal} {Nature Communications}\ }\textbf {\bibinfo {volume} {7}},\
  \bibinfo {pages} {12182} (\bibinfo {year} {2016})}\BibitemShut {NoStop}%
\bibitem [{\citenamefont {Liu}\ \emph {et~al.}(2025)\citenamefont {Liu},
  \citenamefont {Stone}, \citenamefont {Gao}, \citenamefont {Nakamura},
  \citenamefont {Kamazawa}, \citenamefont {Krajewska}, \citenamefont {Walker},
  \citenamefont {Cheng}, \citenamefont {Yu}, \citenamefont {Si}, \citenamefont
  {Dai},\ and\ \citenamefont {Lu}}]{Liu2025}%
  \BibitemOpen
  \bibfield  {author} {\bibinfo {author} {\bibfnamefont {R.}~\bibnamefont
  {Liu}}, \bibinfo {author} {\bibfnamefont {M.~B.}\ \bibnamefont {Stone}},
  \bibinfo {author} {\bibfnamefont {S.}~\bibnamefont {Gao}}, \bibinfo {author}
  {\bibfnamefont {M.}~\bibnamefont {Nakamura}}, \bibinfo {author}
  {\bibfnamefont {K.}~\bibnamefont {Kamazawa}}, \bibinfo {author}
  {\bibfnamefont {A.}~\bibnamefont {Krajewska}}, \bibinfo {author}
  {\bibfnamefont {H.~C.}\ \bibnamefont {Walker}}, \bibinfo {author}
  {\bibfnamefont {P.}~\bibnamefont {Cheng}}, \bibinfo {author} {\bibfnamefont
  {R.}~\bibnamefont {Yu}}, \bibinfo {author} {\bibfnamefont {Q.}~\bibnamefont
  {Si}}, \bibinfo {author} {\bibfnamefont {P.}~\bibnamefont {Dai}},\ and\
  \bibinfo {author} {\bibfnamefont {X.}~\bibnamefont {Lu}},\ }\bibfield
  {title} {\bibinfo {title} {{Spin correlations in the nematic quantum
  disordered state of FeSe}},\ }\href
  {https://doi.org/10.1038/s41467-025-60071-2} {\bibfield  {journal} {\bibinfo
  {journal} {Nature Communications}\ }\textbf {\bibinfo {volume} {16}},\
  \bibinfo {pages} {5212} (\bibinfo {year} {2025})}\BibitemShut {NoStop}%
\bibitem [{\citenamefont {Bristow}\ \emph {et~al.}(2020)\citenamefont
  {Bristow}, \citenamefont {Reiss}, \citenamefont {Haghighirad}, \citenamefont
  {Zajicek}, \citenamefont {Singh}, \citenamefont {Wolf}, \citenamefont {Graf},
  \citenamefont {Knafo}, \citenamefont {McCollam},\ and\ \citenamefont
  {Coldea}}]{Bristow2020}%
  \BibitemOpen
  \bibfield  {author} {\bibinfo {author} {\bibfnamefont {M.}~\bibnamefont
  {Bristow}}, \bibinfo {author} {\bibfnamefont {P.}~\bibnamefont {Reiss}},
  \bibinfo {author} {\bibfnamefont {A.~A.}\ \bibnamefont {Haghighirad}},
  \bibinfo {author} {\bibfnamefont {Z.}~\bibnamefont {Zajicek}}, \bibinfo
  {author} {\bibfnamefont {S.~J.}\ \bibnamefont {Singh}}, \bibinfo {author}
  {\bibfnamefont {T.}~\bibnamefont {Wolf}}, \bibinfo {author} {\bibfnamefont
  {D.}~\bibnamefont {Graf}}, \bibinfo {author} {\bibfnamefont {W.}~\bibnamefont
  {Knafo}}, \bibinfo {author} {\bibfnamefont {A.}~\bibnamefont {McCollam}},\
  and\ \bibinfo {author} {\bibfnamefont {A.~I.}\ \bibnamefont {Coldea}},\
  }\bibfield  {title} {\bibinfo {title} {Anomalous high-magnetic field
  electronic state of the nematic superconductors
  {${\mathrm{FeSe}}_{1\ensuremath{-}x}{\mathrm{S}}_{x}$}},\ }\href
  {https://doi.org/10.1103/PhysRevResearch.2.013309} {\bibfield  {journal}
  {\bibinfo  {journal} {Phys. Rev. Res.}\ }\textbf {\bibinfo {volume} {2}},\
  \bibinfo {pages} {013309} (\bibinfo {year} {2020})}\BibitemShut {NoStop}%
\bibitem [{\citenamefont {Bristow}(2020)}]{Bristow_thesis}%
  \BibitemOpen
  \bibfield  {author} {\bibinfo {author} {\bibfnamefont {M.}~\bibnamefont
  {Bristow}},\ }\emph {\bibinfo {title} {Iron-based superconductors in high
  magnetic fields}},\ \href@noop {} {Ph.D. thesis},\ \bibinfo  {school}
  {University of Oxford} (\bibinfo {year} {2020})\BibitemShut {NoStop}%
\bibitem [{\citenamefont {Zajicek}\ \emph {et~al.}(2026)\citenamefont
  {Zajicek}, \citenamefont {Paulescu}, \citenamefont {Reiss}, \citenamefont
  {Abedin}, \citenamefont {Sun}, \citenamefont {Singh}, \citenamefont
  {Haghighirad},\ and\ \citenamefont {Coldea}}]{Zajicek2026}%
  \BibitemOpen
  \bibfield  {author} {\bibinfo {author} {\bibfnamefont {Z.}~\bibnamefont
  {Zajicek}}, \bibinfo {author} {\bibfnamefont {I.}~\bibnamefont {Paulescu}},
  \bibinfo {author} {\bibfnamefont {P.}~\bibnamefont {Reiss}}, \bibinfo
  {author} {\bibfnamefont {R.~M.}\ \bibnamefont {Abedin}}, \bibinfo {author}
  {\bibfnamefont {K.}~\bibnamefont {Sun}}, \bibinfo {author} {\bibfnamefont
  {S.~J.}\ \bibnamefont {Singh}}, \bibinfo {author} {\bibfnamefont {A.~A.}\
  \bibnamefont {Haghighirad}},\ and\ \bibinfo {author} {\bibfnamefont {A.~I.}\
  \bibnamefont {Coldea}},\ }\bibfield  {title} {\bibinfo {title} {Drastic
  field-induced resistivity upturns as signatures of unconventional magnetism
  in superconducting iron chalcogenides},\ }\href
  {https://doi.org/10.1103/gbfx-x9w1} {\bibfield  {journal} {\bibinfo
  {journal} {Phys. Rev. B}\ }\textbf {\bibinfo {volume} {113}},\ \bibinfo
  {pages} {075135} (\bibinfo {year} {2026})}\BibitemShut {NoStop}%
\bibitem [{\citenamefont {Hanaguri}\ \emph {et~al.}(2018)\citenamefont
  {Hanaguri}, \citenamefont {Iwaya}, \citenamefont {Kohsaka}, \citenamefont
  {Machida}, \citenamefont {Watashige}, \citenamefont {Kasahara}, \citenamefont
  {Shibauchi},\ and\ \citenamefont {Matsuda}}]{Hanaguri2018}%
  \BibitemOpen
  \bibfield  {author} {\bibinfo {author} {\bibfnamefont {T.}~\bibnamefont
  {Hanaguri}}, \bibinfo {author} {\bibfnamefont {K.}~\bibnamefont {Iwaya}},
  \bibinfo {author} {\bibfnamefont {Y.}~\bibnamefont {Kohsaka}}, \bibinfo
  {author} {\bibfnamefont {T.}~\bibnamefont {Machida}}, \bibinfo {author}
  {\bibfnamefont {T.}~\bibnamefont {Watashige}}, \bibinfo {author}
  {\bibfnamefont {S.}~\bibnamefont {Kasahara}}, \bibinfo {author}
  {\bibfnamefont {T.}~\bibnamefont {Shibauchi}},\ and\ \bibinfo {author}
  {\bibfnamefont {Y.}~\bibnamefont {Matsuda}},\ }\bibfield  {title} {\bibinfo
  {title} {{Two distinct superconducting pairing states divided by the nematic
  end point in FeSe$_{1-x}$S$_x$}},\ }\href
  {https://doi.org/10.1126/sciadv.aar6419} {\bibfield  {journal} {\bibinfo
  {journal} {Science Advances}\ }\textbf {\bibinfo {volume} {4}},\ \bibinfo
  {pages} {eaar6419} (\bibinfo {year} {2018})}\BibitemShut {NoStop}%
\bibitem [{\citenamefont {Sun}\ \emph {et~al.}(2016)\citenamefont {Sun},
  \citenamefont {Matsuura}, \citenamefont {Ye}, \citenamefont {Mizukami},
  \citenamefont {Shimozawa}, \citenamefont {Matsubayashi}, \citenamefont
  {Yamashita}, \citenamefont {Watashige}, \citenamefont {Kasahara},
  \citenamefont {Matsuda}, \citenamefont {Yan}, \citenamefont {Sales},
  \citenamefont {Uwatoko}, \citenamefont {Cheng},\ and\ \citenamefont
  {Shibauchi}}]{Sun2016}%
  \BibitemOpen
  \bibfield  {author} {\bibinfo {author} {\bibfnamefont {J.~P.}\ \bibnamefont
  {Sun}}, \bibinfo {author} {\bibfnamefont {K.}~\bibnamefont {Matsuura}},
  \bibinfo {author} {\bibfnamefont {G.~Z.}\ \bibnamefont {Ye}}, \bibinfo
  {author} {\bibfnamefont {Y.}~\bibnamefont {Mizukami}}, \bibinfo {author}
  {\bibfnamefont {M.}~\bibnamefont {Shimozawa}}, \bibinfo {author}
  {\bibfnamefont {K.}~\bibnamefont {Matsubayashi}}, \bibinfo {author}
  {\bibfnamefont {M.}~\bibnamefont {Yamashita}}, \bibinfo {author}
  {\bibfnamefont {T.}~\bibnamefont {Watashige}}, \bibinfo {author}
  {\bibfnamefont {S.}~\bibnamefont {Kasahara}}, \bibinfo {author}
  {\bibfnamefont {Y.}~\bibnamefont {Matsuda}}, \bibinfo {author} {\bibfnamefont
  {J.~Q.}\ \bibnamefont {Yan}}, \bibinfo {author} {\bibfnamefont {B.~C.}\
  \bibnamefont {Sales}}, \bibinfo {author} {\bibfnamefont {Y.}~\bibnamefont
  {Uwatoko}}, \bibinfo {author} {\bibfnamefont {J.~G.}\ \bibnamefont {Cheng}},\
  and\ \bibinfo {author} {\bibfnamefont {T.}~\bibnamefont {Shibauchi}},\
  }\bibfield  {title} {\bibinfo {title} {{Dome-shaped magnetic order competing
  with high-temperature superconductivity at high pressures in FeSe}},\ }\href
  {https://doi.org/10.1038/ncomms12146} {\bibfield  {journal} {\bibinfo
  {journal} {Nature Communications}\ }\textbf {\bibinfo {volume} {7}},\
  \bibinfo {pages} {12146} (\bibinfo {year} {2016})}\BibitemShut {NoStop}%
\bibitem [{\citenamefont {Matsuura}\ \emph {et~al.}(2017)\citenamefont
  {Matsuura}, \citenamefont {Mizukami}, \citenamefont {Arai}, \citenamefont
  {Sugimura}, \citenamefont {Maejima}, \citenamefont {Machida}, \citenamefont
  {Watanuki}, \citenamefont {Fukuda}, \citenamefont {Yajima}, \citenamefont
  {Hiroi}, \citenamefont {Yip}, \citenamefont {Chan}, \citenamefont {Niu},
  \citenamefont {Hosoi}, \citenamefont {Ishida}, \citenamefont {Mukasa},
  \citenamefont {Kasahara}, \citenamefont {Cheng}, \citenamefont {Goh},
  \citenamefont {Matsuda}, \citenamefont {Uwatoko},\ and\ \citenamefont
  {Shibauchi}}]{Matsuura2017}%
  \BibitemOpen
  \bibfield  {author} {\bibinfo {author} {\bibfnamefont {K.}~\bibnamefont
  {Matsuura}}, \bibinfo {author} {\bibfnamefont {Y.}~\bibnamefont {Mizukami}},
  \bibinfo {author} {\bibfnamefont {Y.}~\bibnamefont {Arai}}, \bibinfo {author}
  {\bibfnamefont {Y.}~\bibnamefont {Sugimura}}, \bibinfo {author}
  {\bibfnamefont {N.}~\bibnamefont {Maejima}}, \bibinfo {author} {\bibfnamefont
  {A.}~\bibnamefont {Machida}}, \bibinfo {author} {\bibfnamefont
  {T.}~\bibnamefont {Watanuki}}, \bibinfo {author} {\bibfnamefont
  {T.}~\bibnamefont {Fukuda}}, \bibinfo {author} {\bibfnamefont
  {T.}~\bibnamefont {Yajima}}, \bibinfo {author} {\bibfnamefont
  {Z.}~\bibnamefont {Hiroi}}, \bibinfo {author} {\bibfnamefont {K.~Y.}\
  \bibnamefont {Yip}}, \bibinfo {author} {\bibfnamefont {Y.~C.}\ \bibnamefont
  {Chan}}, \bibinfo {author} {\bibfnamefont {Q.}~\bibnamefont {Niu}}, \bibinfo
  {author} {\bibfnamefont {S.}~\bibnamefont {Hosoi}}, \bibinfo {author}
  {\bibfnamefont {K.}~\bibnamefont {Ishida}}, \bibinfo {author} {\bibfnamefont
  {K.}~\bibnamefont {Mukasa}}, \bibinfo {author} {\bibfnamefont
  {S.}~\bibnamefont {Kasahara}}, \bibinfo {author} {\bibfnamefont {J.-G.}\
  \bibnamefont {Cheng}}, \bibinfo {author} {\bibfnamefont {S.~K.}\ \bibnamefont
  {Goh}}, \bibinfo {author} {\bibfnamefont {Y.}~\bibnamefont {Matsuda}},
  \bibinfo {author} {\bibfnamefont {Y.}~\bibnamefont {Uwatoko}},\ and\ \bibinfo
  {author} {\bibfnamefont {T.}~\bibnamefont {Shibauchi}},\ }\bibfield  {title}
  {\bibinfo {title} {{Maximizing $T_{\rm c}$ by tuning nematicity and magnetism
  in FeSe$_{1-x}$S$_ x$ superconductors}},\ }\href
  {https://doi.org/10.1038/s41467-017-01277-x} {\bibfield  {journal} {\bibinfo
  {journal} {Nat. Comm.}\ }\textbf {\bibinfo {volume} {8}},\ \bibinfo {pages}
  {1143} (\bibinfo {year} {2017})}\BibitemShut {NoStop}%
\bibitem [{\citenamefont {Bendele}\ \emph {et~al.}(2012)\citenamefont
  {Bendele}, \citenamefont {Ichsanow}, \citenamefont {Pashkevich},
  \citenamefont {Keller}, \citenamefont {Str\"assle}, \citenamefont {Gusev},
  \citenamefont {Pomjakushina}, \citenamefont {Conder}, \citenamefont
  {Khasanov},\ and\ \citenamefont {Keller}}]{Bendele2012}%
  \BibitemOpen
  \bibfield  {author} {\bibinfo {author} {\bibfnamefont {M.}~\bibnamefont
  {Bendele}}, \bibinfo {author} {\bibfnamefont {A.}~\bibnamefont {Ichsanow}},
  \bibinfo {author} {\bibfnamefont {Y.}~\bibnamefont {Pashkevich}}, \bibinfo
  {author} {\bibfnamefont {L.}~\bibnamefont {Keller}}, \bibinfo {author}
  {\bibfnamefont {T.}~\bibnamefont {Str\"assle}}, \bibinfo {author}
  {\bibfnamefont {A.}~\bibnamefont {Gusev}}, \bibinfo {author} {\bibfnamefont
  {E.}~\bibnamefont {Pomjakushina}}, \bibinfo {author} {\bibfnamefont
  {K.}~\bibnamefont {Conder}}, \bibinfo {author} {\bibfnamefont
  {R.}~\bibnamefont {Khasanov}},\ and\ \bibinfo {author} {\bibfnamefont
  {H.}~\bibnamefont {Keller}},\ }\bibfield  {title} {\bibinfo {title}
  {Coexistence of superconductivity and magnetism in
  {FeSe${}_{1\ensuremath{-}x}$S${}_{x}$} under pressure},\ }\href
  {https://doi.org/10.1103/PhysRevB.85.064517} {\bibfield  {journal} {\bibinfo
  {journal} {Phys. Rev. B}\ }\textbf {\bibinfo {volume} {85}},\ \bibinfo
  {pages} {064517} (\bibinfo {year} {2012})}\BibitemShut {NoStop}%
\bibitem [{\citenamefont {Kothapalli}\ \emph {et~al.}(2016)\citenamefont
  {Kothapalli}, \citenamefont {B{\"{o}}hmer}, \citenamefont {Jayasekara},
  \citenamefont {Ueland}, \citenamefont {Das}, \citenamefont {Sapkota},
  \citenamefont {Taufour}, \citenamefont {Xiao}, \citenamefont {Alp},
  \citenamefont {Bud'ko}, \citenamefont {Canfield}, \citenamefont {Kreyssig},\
  and\ \citenamefont {Goldman}}]{Kothapalli2016}%
  \BibitemOpen
  \bibfield  {author} {\bibinfo {author} {\bibfnamefont {K.}~\bibnamefont
  {Kothapalli}}, \bibinfo {author} {\bibfnamefont {A.~E.}\ \bibnamefont
  {B{\"{o}}hmer}}, \bibinfo {author} {\bibfnamefont {W.~T.}\ \bibnamefont
  {Jayasekara}}, \bibinfo {author} {\bibfnamefont {B.~G.}\ \bibnamefont
  {Ueland}}, \bibinfo {author} {\bibfnamefont {P.}~\bibnamefont {Das}},
  \bibinfo {author} {\bibfnamefont {A.}~\bibnamefont {Sapkota}}, \bibinfo
  {author} {\bibfnamefont {V.}~\bibnamefont {Taufour}}, \bibinfo {author}
  {\bibfnamefont {Y.}~\bibnamefont {Xiao}}, \bibinfo {author} {\bibfnamefont
  {E.}~\bibnamefont {Alp}}, \bibinfo {author} {\bibfnamefont {S.~L.}\
  \bibnamefont {Bud'ko}}, \bibinfo {author} {\bibfnamefont {P.~C.}\
  \bibnamefont {Canfield}}, \bibinfo {author} {\bibfnamefont {A.}~\bibnamefont
  {Kreyssig}},\ and\ \bibinfo {author} {\bibfnamefont {A.~I.}\ \bibnamefont
  {Goldman}},\ }\bibfield  {title} {\bibinfo {title} {{Strong cooperative
  coupling of pressure-induced magnetic order and nematicity in FeSe}},\ }\href
  {http://dx.doi.org/10.1038/ncomms12728} {\bibfield  {journal} {\bibinfo
  {journal} {Nat. Comm.}\ }\textbf {\bibinfo {volume} {7}},\ \bibinfo {pages}
  {12728} (\bibinfo {year} {2016})}\BibitemShut {NoStop}%
\bibitem [{\citenamefont {Reiss}\ \emph {et~al.}(2024)\citenamefont {Reiss},
  \citenamefont {McCollam}, \citenamefont {Zajicek}, \citenamefont
  {Haghighirad},\ and\ \citenamefont {Coldea}}]{Reiss2024}%
  \BibitemOpen
  \bibfield  {author} {\bibinfo {author} {\bibfnamefont {P.}~\bibnamefont
  {Reiss}}, \bibinfo {author} {\bibfnamefont {A.}~\bibnamefont {McCollam}},
  \bibinfo {author} {\bibfnamefont {Z.}~\bibnamefont {Zajicek}}, \bibinfo
  {author} {\bibfnamefont {A.~A.}\ \bibnamefont {Haghighirad}},\ and\ \bibinfo
  {author} {\bibfnamefont {A.~I.}\ \bibnamefont {Coldea}},\ }\bibfield  {title}
  {\bibinfo {title} {{Collapse of metallicity and high-Tc superconductivity in
  the high-pressure phase of FeSe$_{0.89}$S$_{0.11}$}},\ }\href
  {https://doi.org/10.1038/s41535-024-00677-9} {\bibfield  {journal} {\bibinfo
  {journal} {npj Quantum Materials}\ }\textbf {\bibinfo {volume} {9}},\
  \bibinfo {pages} {73} (\bibinfo {year} {2024})}\BibitemShut {NoStop}%
\bibitem [{\citenamefont {Zajicek}\ \emph
  {et~al.}(2022{\natexlab{a}})\citenamefont {Zajicek}, \citenamefont {Singh},\
  and\ \citenamefont {Coldea}}]{Zajicek2022Cupressure}%
  \BibitemOpen
  \bibfield  {author} {\bibinfo {author} {\bibfnamefont {Z.}~\bibnamefont
  {Zajicek}}, \bibinfo {author} {\bibfnamefont {S.~J.}\ \bibnamefont {Singh}},\
  and\ \bibinfo {author} {\bibfnamefont {A.~I.}\ \bibnamefont {Coldea}},\
  }\bibfield  {title} {\bibinfo {title} {{Robust superconductivity and fragile
  magnetism induced by the strong Cu impurity scattering in the high-pressure
  phase of FeSe}},\ }\href {https://doi.org/10.1103/PhysRevResearch.4.043123}
  {\bibfield  {journal} {\bibinfo  {journal} {Phys. Rev. Res.}\ }\textbf
  {\bibinfo {volume} {4}},\ \bibinfo {pages} {043123} (\bibinfo {year}
  {2022}{\natexlab{a}})}\BibitemShut {NoStop}%
\bibitem [{\citenamefont {Xiang}\ \emph {et~al.}(2017)\citenamefont {Xiang},
  \citenamefont {Kaluarachchi}, \citenamefont {B\"ohmer}, \citenamefont
  {Taufour}, \citenamefont {Tanatar}, \citenamefont {Prozorov}, \citenamefont
  {Bud'ko},\ and\ \citenamefont {Canfield}}]{Xiang2017}%
  \BibitemOpen
  \bibfield  {author} {\bibinfo {author} {\bibfnamefont {L.}~\bibnamefont
  {Xiang}}, \bibinfo {author} {\bibfnamefont {U.~S.}\ \bibnamefont
  {Kaluarachchi}}, \bibinfo {author} {\bibfnamefont {A.~E.}\ \bibnamefont
  {B\"ohmer}}, \bibinfo {author} {\bibfnamefont {V.}~\bibnamefont {Taufour}},
  \bibinfo {author} {\bibfnamefont {M.~A.}\ \bibnamefont {Tanatar}}, \bibinfo
  {author} {\bibfnamefont {R.}~\bibnamefont {Prozorov}}, \bibinfo {author}
  {\bibfnamefont {S.~L.}\ \bibnamefont {Bud'ko}},\ and\ \bibinfo {author}
  {\bibfnamefont {P.~C.}\ \bibnamefont {Canfield}},\ }\bibfield  {title}
  {\bibinfo {title} {{Dome of magnetic order inside the nematic phase of
  sulfur-substituted FeSe under pressure}},\ }\href
  {https://doi.org/10.1103/PhysRevB.96.024511} {\bibfield  {journal} {\bibinfo
  {journal} {Phys. Rev. B}\ }\textbf {\bibinfo {volume} {96}},\ \bibinfo
  {pages} {024511} (\bibinfo {year} {2017})}\BibitemShut {NoStop}%
\bibitem [{\citenamefont {Xie}\ \emph {et~al.}(2021)\citenamefont {Xie},
  \citenamefont {Liu}, \citenamefont {Zhang}, \citenamefont {Wong},
  \citenamefont {Zhou}, \citenamefont {Zhao}, \citenamefont {Wang},
  \citenamefont {Lai},\ and\ \citenamefont {Goh}}]{Xie2021}%
  \BibitemOpen
  \bibfield  {author} {\bibinfo {author} {\bibfnamefont {J.}~\bibnamefont
  {Xie}}, \bibinfo {author} {\bibfnamefont {X.}~\bibnamefont {Liu}}, \bibinfo
  {author} {\bibfnamefont {W.}~\bibnamefont {Zhang}}, \bibinfo {author}
  {\bibfnamefont {S.~M.}\ \bibnamefont {Wong}}, \bibinfo {author}
  {\bibfnamefont {X.}~\bibnamefont {Zhou}}, \bibinfo {author} {\bibfnamefont
  {Y.}~\bibnamefont {Zhao}}, \bibinfo {author} {\bibfnamefont {S.}~\bibnamefont
  {Wang}}, \bibinfo {author} {\bibfnamefont {K.~T.}\ \bibnamefont {Lai}},\ and\
  \bibinfo {author} {\bibfnamefont {S.~K.}\ \bibnamefont {Goh}},\ }\bibfield
  {title} {\bibinfo {title} {{Fragile Pressure-Induced Magnetism in FeSe
  Superconductors with a Thickness Reduction}},\ }\href
  {https://doi.org/10.1021/acs.nanolett.1c03508} {\bibfield  {journal}
  {\bibinfo  {journal} {Nano Letters}\ }\textbf {\bibinfo {volume} {21}},\
  \bibinfo {pages} {9310} (\bibinfo {year} {2021})}\BibitemShut {NoStop}%
\bibitem [{\citenamefont {Zajicek}\ \emph
  {et~al.}(2022{\natexlab{b}})\citenamefont {Zajicek}, \citenamefont {Singh},
  \citenamefont {Jones}, \citenamefont {Reiss}, \citenamefont {Bristow},
  \citenamefont {Martin}, \citenamefont {Gower}, \citenamefont {McCollam},\
  and\ \citenamefont {Coldea}}]{Zajicek2022}%
  \BibitemOpen
  \bibfield  {author} {\bibinfo {author} {\bibfnamefont {Z.}~\bibnamefont
  {Zajicek}}, \bibinfo {author} {\bibfnamefont {S.~J.}\ \bibnamefont {Singh}},
  \bibinfo {author} {\bibfnamefont {H.}~\bibnamefont {Jones}}, \bibinfo
  {author} {\bibfnamefont {P.}~\bibnamefont {Reiss}}, \bibinfo {author}
  {\bibfnamefont {M.}~\bibnamefont {Bristow}}, \bibinfo {author} {\bibfnamefont
  {A.}~\bibnamefont {Martin}}, \bibinfo {author} {\bibfnamefont
  {A.}~\bibnamefont {Gower}}, \bibinfo {author} {\bibfnamefont
  {A.}~\bibnamefont {McCollam}},\ and\ \bibinfo {author} {\bibfnamefont
  {A.~I.}\ \bibnamefont {Coldea}},\ }\bibfield  {title} {\bibinfo {title}
  {Drastic effect of impurity scattering on the electronic and superconducting
  properties of cu-doped fese},\ }\href
  {https://doi.org/10.1103/PhysRevB.105.115130} {\bibfield  {journal} {\bibinfo
   {journal} {Phys. Rev. B}\ }\textbf {\bibinfo {volume} {105}},\ \bibinfo
  {pages} {115130} (\bibinfo {year} {2022}{\natexlab{b}})}\BibitemShut
  {NoStop}%
\bibitem [{\citenamefont {Coldea}\ \emph {et~al.}(2019)\citenamefont {Coldea},
  \citenamefont {Blake}, \citenamefont {Kasahara}, \citenamefont {Haghighirad},
  \citenamefont {Watson}, \citenamefont {Knafo}, \citenamefont {Choi},
  \citenamefont {McCollam}, \citenamefont {Reiss}, \citenamefont {Yamashita},
  \citenamefont {Bruma}, \citenamefont {Speller}, \citenamefont {Matsuda},
  \citenamefont {Wolf}, \citenamefont {Shibauchi},\ and\ \citenamefont
  {Schofield}}]{Coldea2019}%
  \BibitemOpen
  \bibfield  {author} {\bibinfo {author} {\bibfnamefont {A.~I.}\ \bibnamefont
  {Coldea}}, \bibinfo {author} {\bibfnamefont {S.~F.}\ \bibnamefont {Blake}},
  \bibinfo {author} {\bibfnamefont {S.}~\bibnamefont {Kasahara}}, \bibinfo
  {author} {\bibfnamefont {A.~A.}\ \bibnamefont {Haghighirad}}, \bibinfo
  {author} {\bibfnamefont {M.~D.}\ \bibnamefont {Watson}}, \bibinfo {author}
  {\bibfnamefont {W.}~\bibnamefont {Knafo}}, \bibinfo {author} {\bibfnamefont
  {E.~S.}\ \bibnamefont {Choi}}, \bibinfo {author} {\bibfnamefont
  {A.}~\bibnamefont {McCollam}}, \bibinfo {author} {\bibfnamefont
  {P.}~\bibnamefont {Reiss}}, \bibinfo {author} {\bibfnamefont
  {T.}~\bibnamefont {Yamashita}}, \bibinfo {author} {\bibfnamefont
  {M.}~\bibnamefont {Bruma}}, \bibinfo {author} {\bibfnamefont
  {S.}~\bibnamefont {Speller}}, \bibinfo {author} {\bibfnamefont
  {Y.}~\bibnamefont {Matsuda}}, \bibinfo {author} {\bibfnamefont
  {T.}~\bibnamefont {Wolf}}, \bibinfo {author} {\bibfnamefont {T.}~\bibnamefont
  {Shibauchi}},\ and\ \bibinfo {author} {\bibfnamefont {A.~J.}\ \bibnamefont
  {Schofield}},\ }\bibfield  {title} {\bibinfo {title} {{Evolution of the
  low-temperature Fermi surface of superconducting FeSe$_{1-x}$S$_x$ across a
  nematic phase transition}},\ }\href
  {https://doi.org/10.1038/s41535-018-0141-0} {\bibfield  {journal} {\bibinfo
  {journal} {npj Quantum Materials}\ }\textbf {\bibinfo {volume} {4}},\
  \bibinfo {pages} {2} (\bibinfo {year} {2019})}\BibitemShut {NoStop}%
\bibitem [{\citenamefont {Reiss}\ \emph {et~al.}(2020)\citenamefont {Reiss},
  \citenamefont {Graf}, \citenamefont {Haghighirad}, \citenamefont {Knafo},
  \citenamefont {Drigo}, \citenamefont {Bristow}, \citenamefont {Schofield},\
  and\ \citenamefont {Coldea}}]{Reiss2020}%
  \BibitemOpen
  \bibfield  {author} {\bibinfo {author} {\bibfnamefont {P.}~\bibnamefont
  {Reiss}}, \bibinfo {author} {\bibfnamefont {D.}~\bibnamefont {Graf}},
  \bibinfo {author} {\bibfnamefont {A.~A.}\ \bibnamefont {Haghighirad}},
  \bibinfo {author} {\bibfnamefont {W.}~\bibnamefont {Knafo}}, \bibinfo
  {author} {\bibfnamefont {L.}~\bibnamefont {Drigo}}, \bibinfo {author}
  {\bibfnamefont {M.}~\bibnamefont {Bristow}}, \bibinfo {author} {\bibfnamefont
  {A.~J.}\ \bibnamefont {Schofield}},\ and\ \bibinfo {author} {\bibfnamefont
  {A.~I.}\ \bibnamefont {Coldea}},\ }\bibfield  {title} {\bibinfo {title}
  {Quenched nematic criticality and two superconducting domes in an iron-based
  superconductor},\ }\href {https://doi.org/10.1038/s41567-019-0694-2}
  {\bibfield  {journal} {\bibinfo  {journal} {Nature Physics}\ }\textbf
  {\bibinfo {volume} {16}},\ \bibinfo {pages} {89} (\bibinfo {year}
  {2020})}\BibitemShut {NoStop}%
\bibitem [{\citenamefont {Ikeda}\ \emph
  {et~al.}(2018{\natexlab{a}})\citenamefont {Ikeda}, \citenamefont {Worasaran},
  \citenamefont {Palmstrom}, \citenamefont {Straquadine}, \citenamefont
  {Walmsley},\ and\ \citenamefont {Fisher}}]{Ikeda2018}%
  \BibitemOpen
  \bibfield  {author} {\bibinfo {author} {\bibfnamefont {M.~S.}\ \bibnamefont
  {Ikeda}}, \bibinfo {author} {\bibfnamefont {T.}~\bibnamefont {Worasaran}},
  \bibinfo {author} {\bibfnamefont {J.~C.}\ \bibnamefont {Palmstrom}}, \bibinfo
  {author} {\bibfnamefont {J.~A.~W.}\ \bibnamefont {Straquadine}}, \bibinfo
  {author} {\bibfnamefont {P.}~\bibnamefont {Walmsley}},\ and\ \bibinfo
  {author} {\bibfnamefont {I.~R.}\ \bibnamefont {Fisher}},\ }\bibfield  {title}
  {\bibinfo {title} {Symmetric and antisymmetric strain as continuous tuning
  parameters for electronic nematic order},\ }\href
  {https://doi.org/10.1103/PhysRevB.98.245133} {\bibfield  {journal} {\bibinfo
  {journal} {Phys. Rev. B}\ }\textbf {\bibinfo {volume} {98}},\ \bibinfo
  {pages} {245133} (\bibinfo {year} {2018}{\natexlab{a}})}\BibitemShut
  {NoStop}%
\bibitem [{\citenamefont {Lin}\ \emph {et~al.}(2020)\citenamefont {Lin},
  \citenamefont {Campbell}, \citenamefont {Ran}, \citenamefont {Liu},
  \citenamefont {Kim}, \citenamefont {Nevidomskyy}, \citenamefont {Graf},
  \citenamefont {Butch},\ and\ \citenamefont {Paglione}}]{Lin2020}%
  \BibitemOpen
  \bibfield  {author} {\bibinfo {author} {\bibfnamefont {W.-C.}\ \bibnamefont
  {Lin}}, \bibinfo {author} {\bibfnamefont {D.~J.}\ \bibnamefont {Campbell}},
  \bibinfo {author} {\bibfnamefont {S.}~\bibnamefont {Ran}}, \bibinfo {author}
  {\bibfnamefont {I.-L.}\ \bibnamefont {Liu}}, \bibinfo {author} {\bibfnamefont
  {H.}~\bibnamefont {Kim}}, \bibinfo {author} {\bibfnamefont {A.~H.}\
  \bibnamefont {Nevidomskyy}}, \bibinfo {author} {\bibfnamefont
  {D.}~\bibnamefont {Graf}}, \bibinfo {author} {\bibfnamefont {N.~P.}\
  \bibnamefont {Butch}},\ and\ \bibinfo {author} {\bibfnamefont
  {J.}~\bibnamefont {Paglione}},\ }\bibfield  {title} {\bibinfo {title} {Tuning
  magnetic confinement of spin-triplet superconductivity},\ }\href
  {https://doi.org/10.1038/s41535-020-00270-w} {\bibfield  {journal} {\bibinfo
  {journal} {npj Quantum Materials}\ }\textbf {\bibinfo {volume} {5}},\
  \bibinfo {pages} {68} (\bibinfo {year} {2020})}\BibitemShut {NoStop}%
\bibitem [{\citenamefont {Pearce}\ \emph {et~al.}(2024)\citenamefont {Pearce},
  \citenamefont {Kaib}, \citenamefont {Ma}, \citenamefont {Ni}, \citenamefont
  {Cava}, \citenamefont {Valent\'{\i}}, \citenamefont {Coldea},\ and\
  \citenamefont {Coldea}}]{Pearce2024}%
  \BibitemOpen
  \bibfield  {author} {\bibinfo {author} {\bibfnamefont {J.~S.}\ \bibnamefont
  {Pearce}}, \bibinfo {author} {\bibfnamefont {D.~A.~S.}\ \bibnamefont {Kaib}},
  \bibinfo {author} {\bibfnamefont {Z.}~\bibnamefont {Ma}}, \bibinfo {author}
  {\bibfnamefont {D.}~\bibnamefont {Ni}}, \bibinfo {author} {\bibfnamefont
  {R.~J.}\ \bibnamefont {Cava}}, \bibinfo {author} {\bibfnamefont
  {R.}~\bibnamefont {Valent\'{\i}}}, \bibinfo {author} {\bibfnamefont
  {R.}~\bibnamefont {Coldea}},\ and\ \bibinfo {author} {\bibfnamefont {A.~I.}\
  \bibnamefont {Coldea}},\ }\bibfield  {title} {\bibinfo {title} {{Anisotropy
  of the zigzag order in the Kitaev honeycomb magnet
  $\ensuremath{\alpha}\text{\ensuremath{-}}{\mathrm{RuBr}}_{3}$}},\ }\href
  {https://doi.org/10.1103/PhysRevB.110.214404} {\bibfield  {journal} {\bibinfo
   {journal} {Phys. Rev. B}\ }\textbf {\bibinfo {volume} {110}},\ \bibinfo
  {pages} {214404} (\bibinfo {year} {2024})}\BibitemShut {NoStop}%
\bibitem [{\citenamefont {Kasahara}\ \emph {et~al.}(2012)\citenamefont
  {Kasahara}, \citenamefont {Shi}, \citenamefont {Hashimoto}, \citenamefont
  {Tonegawa}, \citenamefont {Mizukami}, \citenamefont {Shibauchi},
  \citenamefont {Sugimoto}, \citenamefont {Fukuda}, \citenamefont {Terashima},
  \citenamefont {Nevidomskyy},\ and\ \citenamefont {Matsuda}}]{Kasahara2012}%
  \BibitemOpen
  \bibfield  {author} {\bibinfo {author} {\bibfnamefont {S.}~\bibnamefont
  {Kasahara}}, \bibinfo {author} {\bibfnamefont {H.~J.}\ \bibnamefont {Shi}},
  \bibinfo {author} {\bibfnamefont {K.}~\bibnamefont {Hashimoto}}, \bibinfo
  {author} {\bibfnamefont {S.}~\bibnamefont {Tonegawa}}, \bibinfo {author}
  {\bibfnamefont {Y.}~\bibnamefont {Mizukami}}, \bibinfo {author}
  {\bibfnamefont {T.}~\bibnamefont {Shibauchi}}, \bibinfo {author}
  {\bibfnamefont {K.}~\bibnamefont {Sugimoto}}, \bibinfo {author}
  {\bibfnamefont {T.}~\bibnamefont {Fukuda}}, \bibinfo {author} {\bibfnamefont
  {T.}~\bibnamefont {Terashima}}, \bibinfo {author} {\bibfnamefont {A.~H.}\
  \bibnamefont {Nevidomskyy}},\ and\ \bibinfo {author} {\bibfnamefont
  {Y.}~\bibnamefont {Matsuda}},\ }\bibfield  {title} {\bibinfo {title}
  {{Electronic nematicity above the structural and superconducting transition
  in BaFe$_2$(As$_{1-x}$P$_x$)$_2$ }},\ }\href
  {https://doi.org/10.1038/nature11178} {\bibfield  {journal} {\bibinfo
  {journal} {Nature}\ }\textbf {\bibinfo {volume} {486}},\ \bibinfo {pages}
  {382} (\bibinfo {year} {2012})}\BibitemShut {NoStop}%
\bibitem [{\citenamefont {Terashima}\ \emph {et~al.}(2016)\citenamefont
  {Terashima}, \citenamefont {Kikugawa}, \citenamefont {Kiswandhi},
  \citenamefont {Graf}, \citenamefont {Choi}, \citenamefont {Brooks},
  \citenamefont {Kasahara}, \citenamefont {Watashige}, \citenamefont {Matsuda},
  \citenamefont {Shibauchi}, \citenamefont {Wolf}, \citenamefont {B\"ohmer},
  \citenamefont {Hardy}, \citenamefont {Meingast}, \citenamefont
  {L\"ohneysen},\ and\ \citenamefont {Uji}}]{Terashima2015}%
  \BibitemOpen
  \bibfield  {author} {\bibinfo {author} {\bibfnamefont {T.}~\bibnamefont
  {Terashima}}, \bibinfo {author} {\bibfnamefont {N.}~\bibnamefont {Kikugawa}},
  \bibinfo {author} {\bibfnamefont {A.}~\bibnamefont {Kiswandhi}}, \bibinfo
  {author} {\bibfnamefont {D.}~\bibnamefont {Graf}}, \bibinfo {author}
  {\bibfnamefont {E.-S.}\ \bibnamefont {Choi}}, \bibinfo {author}
  {\bibfnamefont {J.~S.}\ \bibnamefont {Brooks}}, \bibinfo {author}
  {\bibfnamefont {S.}~\bibnamefont {Kasahara}}, \bibinfo {author}
  {\bibfnamefont {T.}~\bibnamefont {Watashige}}, \bibinfo {author}
  {\bibfnamefont {Y.}~\bibnamefont {Matsuda}}, \bibinfo {author} {\bibfnamefont
  {T.}~\bibnamefont {Shibauchi}}, \bibinfo {author} {\bibfnamefont
  {T.}~\bibnamefont {Wolf}}, \bibinfo {author} {\bibfnamefont {A.~E.}\
  \bibnamefont {B\"ohmer}}, \bibinfo {author} {\bibfnamefont {F.}~\bibnamefont
  {Hardy}}, \bibinfo {author} {\bibfnamefont {C.}~\bibnamefont {Meingast}},
  \bibinfo {author} {\bibfnamefont {H.~v.}\ \bibnamefont {L\"ohneysen}},\ and\
  \bibinfo {author} {\bibfnamefont {S.}~\bibnamefont {Uji}},\ }\bibfield
  {title} {\bibinfo {title} {Fermi surface reconstruction in {FeSe} under high
  pressure},\ }\href {https://doi.org/10.1103/PhysRevB.93.094505} {\bibfield
  {journal} {\bibinfo  {journal} {Phys. Rev. B}\ }\textbf {\bibinfo {volume}
  {93}},\ \bibinfo {pages} {094505} (\bibinfo {year} {2016})}\BibitemShut
  {NoStop}%
\bibitem [{\citenamefont {Lake}\ \emph {et~al.}(2002)\citenamefont {Lake},
  \citenamefont {R{\o}nnow}, \citenamefont {Christensen}, \citenamefont
  {Aeppli}, \citenamefont {Lefmann}, \citenamefont {McMorrow}, \citenamefont
  {Vorderwisch}, \citenamefont {Smeibidl}, \citenamefont {Mangkorntong},
  \citenamefont {Sasagawa}, \citenamefont {Nohara}, \citenamefont {Takagi},\
  and\ \citenamefont {Mason}}]{Lake2002Nature}%
  \BibitemOpen
  \bibfield  {author} {\bibinfo {author} {\bibfnamefont {B.}~\bibnamefont
  {Lake}}, \bibinfo {author} {\bibfnamefont {H.~M.}\ \bibnamefont {R{\o}nnow}},
  \bibinfo {author} {\bibfnamefont {N.~B.}\ \bibnamefont {Christensen}},
  \bibinfo {author} {\bibfnamefont {G.}~\bibnamefont {Aeppli}}, \bibinfo
  {author} {\bibfnamefont {K.}~\bibnamefont {Lefmann}}, \bibinfo {author}
  {\bibfnamefont {D.~F.}\ \bibnamefont {McMorrow}}, \bibinfo {author}
  {\bibfnamefont {P.}~\bibnamefont {Vorderwisch}}, \bibinfo {author}
  {\bibfnamefont {P.}~\bibnamefont {Smeibidl}}, \bibinfo {author}
  {\bibfnamefont {N.}~\bibnamefont {Mangkorntong}}, \bibinfo {author}
  {\bibfnamefont {T.}~\bibnamefont {Sasagawa}}, \bibinfo {author}
  {\bibfnamefont {M.}~\bibnamefont {Nohara}}, \bibinfo {author} {\bibfnamefont
  {H.}~\bibnamefont {Takagi}},\ and\ \bibinfo {author} {\bibfnamefont {T.~E.}\
  \bibnamefont {Mason}},\ }\bibfield  {title} {\bibinfo {title}
  {Antiferromagnetic order induced by an applied magnetic field in a
  high-temperature superconductor},\ }\href {https://doi.org/10.1038/415299a}
  {\bibfield  {journal} {\bibinfo  {journal} {Nature}\ }\textbf {\bibinfo
  {volume} {415}},\ \bibinfo {pages} {299} (\bibinfo {year}
  {2002})}\BibitemShut {NoStop}%
\bibitem [{\citenamefont {Terashima}\ \emph {et~al.}(2014)\citenamefont
  {Terashima}, \citenamefont {Kikugawa}, \citenamefont {Kiswandhi},
  \citenamefont {Choi}, \citenamefont {Brooks}, \citenamefont {Kasahara},
  \citenamefont {Watashige}, \citenamefont {Ikeda}, \citenamefont {Shibauchi},
  \citenamefont {Matsuda}, \citenamefont {Wolf}, \citenamefont {B\"ohmer},
  \citenamefont {Hardy}, \citenamefont {Meingast}, \citenamefont {L\"ohneysen},
  \citenamefont {Suzuki}, \citenamefont {Arita},\ and\ \citenamefont
  {Uji}}]{Terashima2014}%
  \BibitemOpen
  \bibfield  {author} {\bibinfo {author} {\bibfnamefont {T.}~\bibnamefont
  {Terashima}}, \bibinfo {author} {\bibfnamefont {N.}~\bibnamefont {Kikugawa}},
  \bibinfo {author} {\bibfnamefont {A.}~\bibnamefont {Kiswandhi}}, \bibinfo
  {author} {\bibfnamefont {E.-S.}\ \bibnamefont {Choi}}, \bibinfo {author}
  {\bibfnamefont {J.~S.}\ \bibnamefont {Brooks}}, \bibinfo {author}
  {\bibfnamefont {S.}~\bibnamefont {Kasahara}}, \bibinfo {author}
  {\bibfnamefont {T.}~\bibnamefont {Watashige}}, \bibinfo {author}
  {\bibfnamefont {H.}~\bibnamefont {Ikeda}}, \bibinfo {author} {\bibfnamefont
  {T.}~\bibnamefont {Shibauchi}}, \bibinfo {author} {\bibfnamefont
  {Y.}~\bibnamefont {Matsuda}}, \bibinfo {author} {\bibfnamefont
  {T.}~\bibnamefont {Wolf}}, \bibinfo {author} {\bibfnamefont {A.~E.}\
  \bibnamefont {B\"ohmer}}, \bibinfo {author} {\bibfnamefont {F.}~\bibnamefont
  {Hardy}}, \bibinfo {author} {\bibfnamefont {C.}~\bibnamefont {Meingast}},
  \bibinfo {author} {\bibfnamefont {H.~v.}\ \bibnamefont {L\"ohneysen}},
  \bibinfo {author} {\bibfnamefont {M.-T.}\ \bibnamefont {Suzuki}}, \bibinfo
  {author} {\bibfnamefont {R.}~\bibnamefont {Arita}},\ and\ \bibinfo {author}
  {\bibfnamefont {S.}~\bibnamefont {Uji}},\ }\bibfield  {title} {\bibinfo
  {title} {{Anomalous Fermi surface in FeSe seen by Shubnikov--de Haas
  oscillation measurements}},\ }\href
  {https://doi.org/10.1103/PhysRevB.90.144517} {\bibfield  {journal} {\bibinfo
  {journal} {Phys. Rev. B}\ }\textbf {\bibinfo {volume} {90}},\ \bibinfo
  {pages} {144517} (\bibinfo {year} {2014})}\BibitemShut {NoStop}%
\bibitem [{\citenamefont {Terashima}\ \emph {et~al.}(2019)\citenamefont
  {Terashima}, \citenamefont {Kikugawa}, \citenamefont {Graf}, \citenamefont
  {Hirose}, \citenamefont {Uji}, \citenamefont {Matsushita}, \citenamefont
  {Lin}, \citenamefont {Zhu}, \citenamefont {Wen}, \citenamefont {Nomoto},
  \citenamefont {Suzuki},\ and\ \citenamefont {Ikeda}}]{Terashima2019}%
  \BibitemOpen
  \bibfield  {author} {\bibinfo {author} {\bibfnamefont {T.}~\bibnamefont
  {Terashima}}, \bibinfo {author} {\bibfnamefont {N.}~\bibnamefont {Kikugawa}},
  \bibinfo {author} {\bibfnamefont {D.}~\bibnamefont {Graf}}, \bibinfo {author}
  {\bibfnamefont {H.~T.}\ \bibnamefont {Hirose}}, \bibinfo {author}
  {\bibfnamefont {S.}~\bibnamefont {Uji}}, \bibinfo {author} {\bibfnamefont
  {Y.}~\bibnamefont {Matsushita}}, \bibinfo {author} {\bibfnamefont
  {H.}~\bibnamefont {Lin}}, \bibinfo {author} {\bibfnamefont {X.}~\bibnamefont
  {Zhu}}, \bibinfo {author} {\bibfnamefont {H.-H.}\ \bibnamefont {Wen}},
  \bibinfo {author} {\bibfnamefont {T.}~\bibnamefont {Nomoto}}, \bibinfo
  {author} {\bibfnamefont {K.}~\bibnamefont {Suzuki}},\ and\ \bibinfo {author}
  {\bibfnamefont {H.}~\bibnamefont {Ikeda}},\ }\bibfield  {title} {\bibinfo
  {title} {{Accurate determination of the Fermi surface of tetragonal FeS via
  quantum oscillation measurements and quasiparticle self-consistent GW
  calculations}},\ }\href {https://doi.org/10.1103/PhysRevB.99.134501}
  {\bibfield  {journal} {\bibinfo  {journal} {Phys. Rev. B}\ }\textbf {\bibinfo
  {volume} {99}},\ \bibinfo {pages} {134501} (\bibinfo {year}
  {2019})}\BibitemShut {NoStop}%
\bibitem [{\citenamefont {Korshunov}\ \emph {et~al.}(2009)\citenamefont
  {Korshunov}, \citenamefont {Eremin}, \citenamefont {Efremov}, \citenamefont
  {Maslov},\ and\ \citenamefont {Chubukov}}]{Korshunov2009}%
  \BibitemOpen
  \bibfield  {author} {\bibinfo {author} {\bibfnamefont {M.~M.}\ \bibnamefont
  {Korshunov}}, \bibinfo {author} {\bibfnamefont {I.}~\bibnamefont {Eremin}},
  \bibinfo {author} {\bibfnamefont {D.~V.}\ \bibnamefont {Efremov}}, \bibinfo
  {author} {\bibfnamefont {D.~L.}\ \bibnamefont {Maslov}},\ and\ \bibinfo
  {author} {\bibfnamefont {A.~V.}\ \bibnamefont {Chubukov}},\ }\bibfield
  {title} {\bibinfo {title} {{Nonanalytic Spin Susceptibility of a Fermi
  Liquid: The Case of Fe-Based Pnictides}},\ }\href
  {https://doi.org/10.1103/PhysRevLett.102.236403} {\bibfield  {journal}
  {\bibinfo  {journal} {Phys. Rev. Lett.}\ }\textbf {\bibinfo {volume} {102}},\
  \bibinfo {pages} {236403} (\bibinfo {year} {2009})}\BibitemShut {NoStop}%
\bibitem [{\citenamefont {Ok}\ \emph {et~al.}(2020)\citenamefont {Ok},
  \citenamefont {Kwon}, \citenamefont {Kohama}, \citenamefont {You},
  \citenamefont {Park}, \citenamefont {Kim}, \citenamefont {Jo}, \citenamefont
  {Choi}, \citenamefont {Kindo}, \citenamefont {Kang}, \citenamefont {Kim},
  \citenamefont {Moon}, \citenamefont {Gurevich},\ and\ \citenamefont
  {Kim}}]{Ok2020}%
  \BibitemOpen
  \bibfield  {author} {\bibinfo {author} {\bibfnamefont {J.~M.}\ \bibnamefont
  {Ok}}, \bibinfo {author} {\bibfnamefont {C.~I.}\ \bibnamefont {Kwon}},
  \bibinfo {author} {\bibfnamefont {Y.}~\bibnamefont {Kohama}}, \bibinfo
  {author} {\bibfnamefont {J.~S.}\ \bibnamefont {You}}, \bibinfo {author}
  {\bibfnamefont {S.~K.}\ \bibnamefont {Park}}, \bibinfo {author}
  {\bibfnamefont {J.-h.}\ \bibnamefont {Kim}}, \bibinfo {author} {\bibfnamefont
  {Y.~J.}\ \bibnamefont {Jo}}, \bibinfo {author} {\bibfnamefont {E.~S.}\
  \bibnamefont {Choi}}, \bibinfo {author} {\bibfnamefont {K.}~\bibnamefont
  {Kindo}}, \bibinfo {author} {\bibfnamefont {W.}~\bibnamefont {Kang}},
  \bibinfo {author} {\bibfnamefont {K.-S.}\ \bibnamefont {Kim}}, \bibinfo
  {author} {\bibfnamefont {E.~G.}\ \bibnamefont {Moon}}, \bibinfo {author}
  {\bibfnamefont {A.}~\bibnamefont {Gurevich}},\ and\ \bibinfo {author}
  {\bibfnamefont {J.~S.}\ \bibnamefont {Kim}},\ }\bibfield  {title} {\bibinfo
  {title} {Observation of in-plane magnetic field induced phase transitions in
  fese},\ }\href {https://doi.org/10.1103/PhysRevB.101.224509} {\bibfield
  {journal} {\bibinfo  {journal} {Phys. Rev. B}\ }\textbf {\bibinfo {volume}
  {101}},\ \bibinfo {pages} {224509} (\bibinfo {year} {2020})}\BibitemShut
  {NoStop}%
\bibitem [{\citenamefont {Farrar}\ \emph {et~al.}(2020)\citenamefont {Farrar},
  \citenamefont {Bristow}, \citenamefont {Haghighirad}, \citenamefont
  {McCollam}, \citenamefont {Bending},\ and\ \citenamefont
  {Coldea}}]{Farrar2020}%
  \BibitemOpen
  \bibfield  {author} {\bibinfo {author} {\bibfnamefont {L.~S.}\ \bibnamefont
  {Farrar}}, \bibinfo {author} {\bibfnamefont {M.}~\bibnamefont {Bristow}},
  \bibinfo {author} {\bibfnamefont {A.~A.}\ \bibnamefont {Haghighirad}},
  \bibinfo {author} {\bibfnamefont {A.}~\bibnamefont {McCollam}}, \bibinfo
  {author} {\bibfnamefont {S.~J.}\ \bibnamefont {Bending}},\ and\ \bibinfo
  {author} {\bibfnamefont {A.~I.}\ \bibnamefont {Coldea}},\ }\bibfield  {title}
  {\bibinfo {title} {{Suppression of superconductivity and enhanced critical
  field anisotropy in thin flakes of FeSe}},\ }\href
  {https://doi.org/10.1038/s41535-020-0227-3} {\bibfield  {journal} {\bibinfo
  {journal} {npj Quantum Materials}\ }\textbf {\bibinfo {volume} {5}},\
  \bibinfo {pages} {29} (\bibinfo {year} {2020})}\BibitemShut {NoStop}%
\bibitem [{\citenamefont {Morfoot}(2025)}]{Morfoot2025}%
  \BibitemOpen
  \bibfield  {author} {\bibinfo {author} {\bibfnamefont {A.~B.}\ \bibnamefont
  {Morfoot}},\ }\bibfield  {title} {\bibinfo {title} {{PhD Thesis}},\
  }\href@noop {} {\bibfield  {journal} {\bibinfo  {journal} {University of
  Oxford}\ } (\bibinfo {year} {2025})}\BibitemShut {NoStop}%
\bibitem [{\citenamefont {Zajicek}\ \emph {et~al.}(2024)\citenamefont
  {Zajicek}, \citenamefont {Reiss}, \citenamefont {Graf}, \citenamefont
  {Prentice}, \citenamefont {Sadki}, \citenamefont {Haghighirad},\ and\
  \citenamefont {Coldea}}]{Zajicek2024}%
  \BibitemOpen
  \bibfield  {author} {\bibinfo {author} {\bibfnamefont {Z.}~\bibnamefont
  {Zajicek}}, \bibinfo {author} {\bibfnamefont {P.}~\bibnamefont {Reiss}},
  \bibinfo {author} {\bibfnamefont {D.}~\bibnamefont {Graf}}, \bibinfo {author}
  {\bibfnamefont {J.~C.}\ \bibnamefont {Prentice}}, \bibinfo {author}
  {\bibfnamefont {Y.}~\bibnamefont {Sadki}}, \bibinfo {author} {\bibfnamefont
  {A.~A.}\ \bibnamefont {Haghighirad}},\ and\ \bibinfo {author} {\bibfnamefont
  {A.~I.}\ \bibnamefont {Coldea}},\ }\bibfield  {title} {\bibinfo {title}
  {{Unveiling the quasiparticle behaviour in the pressure-induced
  high-$\textit{T}_{c}$ phase of an iron-chalcogenide superconductor}},\
  }\bibfield  {journal} {\bibinfo  {journal} {npj Quantum Materials}\ }\textbf
  {\bibinfo {volume} {9}},\ \href {https://doi.org/10.1038/s41535-024-00663-1}
  {10.1038/s41535-024-00663-1} (\bibinfo {year} {2024})\BibitemShut {NoStop}%
\bibitem [{\citenamefont {Moon}\ and\ \citenamefont {Choi}(2010)}]{Moon2010}%
  \BibitemOpen
  \bibfield  {author} {\bibinfo {author} {\bibfnamefont {C.-Y.}\ \bibnamefont
  {Moon}}\ and\ \bibinfo {author} {\bibfnamefont {H.~J.}\ \bibnamefont
  {Choi}},\ }\bibfield  {title} {\bibinfo {title} {{Chalcogen-Height Dependent
  Magnetic Interactions and Magnetic Order Switching in
  ${\mathrm{FeSe}}_{x}{\mathrm{Te}}_{1\ensuremath{-}x}$}},\ }\href
  {https://doi.org/10.1103/PhysRevLett.104.057003} {\bibfield  {journal}
  {\bibinfo  {journal} {Phys. Rev. Lett.}\ }\textbf {\bibinfo {volume} {104}},\
  \bibinfo {pages} {057003} (\bibinfo {year} {2010})}\BibitemShut {NoStop}%
\bibitem [{\citenamefont {Yamakawa}\ and\ \citenamefont
  {Kontani}(2017)}]{Yamakaua2017}%
  \BibitemOpen
  \bibfield  {author} {\bibinfo {author} {\bibfnamefont {Y.}~\bibnamefont
  {Yamakawa}}\ and\ \bibinfo {author} {\bibfnamefont {H.}~\bibnamefont
  {Kontani}},\ }\bibfield  {title} {\bibinfo {title} {{Nematicity, magnetism,
  and superconductivity in FeSe under pressure: Unified explanation based on
  the self-consistent vertex correction theory}},\ }\href
  {https://doi.org/10.1103/PhysRevB.96.144509} {\bibfield  {journal} {\bibinfo
  {journal} {Phys. Rev. B}\ }\textbf {\bibinfo {volume} {96}},\ \bibinfo
  {pages} {144509} (\bibinfo {year} {2017})}\BibitemShut {NoStop}%
\bibitem [{\citenamefont {Dhaka}\ \emph {et~al.}(2011)\citenamefont {Dhaka},
  \citenamefont {Liu}, \citenamefont {Fernandes}, \citenamefont {Jiang},
  \citenamefont {Strehlow}, \citenamefont {Kondo}, \citenamefont {Thaler},
  \citenamefont {Schmalian}, \citenamefont {Bud'ko}, \citenamefont {Canfield},\
  and\ \citenamefont {Kaminski}}]{Dhaka2011}%
  \BibitemOpen
  \bibfield  {author} {\bibinfo {author} {\bibfnamefont {R.~S.}\ \bibnamefont
  {Dhaka}}, \bibinfo {author} {\bibfnamefont {C.}~\bibnamefont {Liu}}, \bibinfo
  {author} {\bibfnamefont {R.~M.}\ \bibnamefont {Fernandes}}, \bibinfo {author}
  {\bibfnamefont {R.}~\bibnamefont {Jiang}}, \bibinfo {author} {\bibfnamefont
  {C.~P.}\ \bibnamefont {Strehlow}}, \bibinfo {author} {\bibfnamefont
  {T.}~\bibnamefont {Kondo}}, \bibinfo {author} {\bibfnamefont
  {A.}~\bibnamefont {Thaler}}, \bibinfo {author} {\bibfnamefont
  {J.}~\bibnamefont {Schmalian}}, \bibinfo {author} {\bibfnamefont {S.~L.}\
  \bibnamefont {Bud'ko}}, \bibinfo {author} {\bibfnamefont {P.~C.}\
  \bibnamefont {Canfield}},\ and\ \bibinfo {author} {\bibfnamefont
  {A.}~\bibnamefont {Kaminski}},\ }\bibfield  {title} {\bibinfo {title} {{What
  Controls the Phase Diagram and Superconductivity in Ru-Substituted
  ${\mathrm{BaFe}}_{2}{\mathrm{As}}_{2}$?}},\ }\href
  {https://doi.org/10.1103/PhysRevLett.107.267002} {\bibfield  {journal}
  {\bibinfo  {journal} {Phys. Rev. Lett.}\ }\textbf {\bibinfo {volume} {107}},\
  \bibinfo {pages} {267002} (\bibinfo {year} {2011})}\BibitemShut {NoStop}%
\bibitem [{\citenamefont {Ishida}\ \emph {et~al.}(2013)\citenamefont {Ishida},
  \citenamefont {Nakajima}, \citenamefont {Liang}, \citenamefont {Kihou},
  \citenamefont {Lee}, \citenamefont {Iyo}, \citenamefont {Eisaki},
  \citenamefont {Kakeshita}, \citenamefont {Tomioka}, \citenamefont {Ito},\
  and\ \citenamefont {Uchida}}]{Ishida2013}%
  \BibitemOpen
  \bibfield  {author} {\bibinfo {author} {\bibfnamefont {S.}~\bibnamefont
  {Ishida}}, \bibinfo {author} {\bibfnamefont {M.}~\bibnamefont {Nakajima}},
  \bibinfo {author} {\bibfnamefont {T.}~\bibnamefont {Liang}}, \bibinfo
  {author} {\bibfnamefont {K.}~\bibnamefont {Kihou}}, \bibinfo {author}
  {\bibfnamefont {C.-H.}\ \bibnamefont {Lee}}, \bibinfo {author} {\bibfnamefont
  {A.}~\bibnamefont {Iyo}}, \bibinfo {author} {\bibfnamefont {H.}~\bibnamefont
  {Eisaki}}, \bibinfo {author} {\bibfnamefont {T.}~\bibnamefont {Kakeshita}},
  \bibinfo {author} {\bibfnamefont {Y.}~\bibnamefont {Tomioka}}, \bibinfo
  {author} {\bibfnamefont {T.}~\bibnamefont {Ito}},\ and\ \bibinfo {author}
  {\bibfnamefont {S.-i.}\ \bibnamefont {Uchida}},\ }\bibfield  {title}
  {\bibinfo {title} {{Effect of Doping on the Magnetostructural Ordered Phase
  of Iron Arsenides: A Comparative Study of the Resistivity Anisotropy in Doped
  BaFe$_2$As$_2$ with Doping into Three Different Sites}},\ }\href
  {https://doi.org/10.1021/ja311174e} {\bibfield  {journal} {\bibinfo
  {journal} {Journal of the American Chemical Society}\ }\textbf {\bibinfo
  {volume} {135}},\ \bibinfo {pages} {3158} (\bibinfo {year}
  {2013})}\BibitemShut {NoStop}%
\bibitem [{\citenamefont {Wiecki}\ \emph {et~al.}(2017)\citenamefont {Wiecki},
  \citenamefont {Nandi}, \citenamefont {B\"ohmer}, \citenamefont {Bud'ko},
  \citenamefont {Canfield},\ and\ \citenamefont {Furukawa}}]{Wiecki2017}%
  \BibitemOpen
  \bibfield  {author} {\bibinfo {author} {\bibfnamefont {P.}~\bibnamefont
  {Wiecki}}, \bibinfo {author} {\bibfnamefont {M.}~\bibnamefont {Nandi}},
  \bibinfo {author} {\bibfnamefont {A.~E.}\ \bibnamefont {B\"ohmer}}, \bibinfo
  {author} {\bibfnamefont {S.~L.}\ \bibnamefont {Bud'ko}}, \bibinfo {author}
  {\bibfnamefont {P.~C.}\ \bibnamefont {Canfield}},\ and\ \bibinfo {author}
  {\bibfnamefont {Y.}~\bibnamefont {Furukawa}},\ }\bibfield  {title} {\bibinfo
  {title} {{NMR evidence for static local nematicity and its cooperative
  interplay with low-energy magnetic fluctuations in FeSe under pressure}},\
  }\href {https://doi.org/10.1103/PhysRevB.96.180502} {\bibfield  {journal}
  {\bibinfo  {journal} {Phys. Rev. B}\ }\textbf {\bibinfo {volume} {96}},\
  \bibinfo {pages} {180502} (\bibinfo {year} {2017})}\BibitemShut {NoStop}%
\bibitem [{\citenamefont {Fernandes}\ and\ \citenamefont
  {Schmalian}(2010{\natexlab{b}})}]{Fernandes2010optical}%
  \BibitemOpen
  \bibfield  {author} {\bibinfo {author} {\bibfnamefont {R.~M.}\ \bibnamefont
  {Fernandes}}\ and\ \bibinfo {author} {\bibfnamefont {J.}~\bibnamefont
  {Schmalian}},\ }\bibfield  {title} {\bibinfo {title} {{Transfer of optical
  spectral weight in magnetically ordered superconductors}},\ }\href
  {https://doi.org/10.1103/PhysRevB.82.014520} {\bibfield  {journal} {\bibinfo
  {journal} {Phys. Rev. B}\ }\textbf {\bibinfo {volume} {82}},\ \bibinfo
  {pages} {014520} (\bibinfo {year} {2010}{\natexlab{b}})}\BibitemShut
  {NoStop}%
\bibitem [{\citenamefont {Cao}\ \emph {et~al.}(2023)\citenamefont {Cao},
  \citenamefont {Setty}, \citenamefont {Fanfarillo}, \citenamefont {Kreisel},\
  and\ \citenamefont {Hirschfeld}}]{Peter2023}%
  \BibitemOpen
  \bibfield  {author} {\bibinfo {author} {\bibfnamefont {Y.}~\bibnamefont
  {Cao}}, \bibinfo {author} {\bibfnamefont {C.}~\bibnamefont {Setty}}, \bibinfo
  {author} {\bibfnamefont {L.}~\bibnamefont {Fanfarillo}}, \bibinfo {author}
  {\bibfnamefont {A.}~\bibnamefont {Kreisel}},\ and\ \bibinfo {author}
  {\bibfnamefont {P.~J.}\ \bibnamefont {Hirschfeld}},\ }\bibfield  {title}
  {\bibinfo {title} {Microscopic origin of ultranodal superconducting states in
  spin-{$\frac{1}{2}$} systems},\ }\href
  {https://doi.org/10.1103/PhysRevB.108.224506} {\bibfield  {journal} {\bibinfo
   {journal} {Phys. Rev. B}\ }\textbf {\bibinfo {volume} {108}},\ \bibinfo
  {pages} {224506} (\bibinfo {year} {2023})}\BibitemShut {NoStop}%
\bibitem [{\citenamefont {Wu}\ \emph {et~al.}(2024)\citenamefont {Wu},
  \citenamefont {Amin}, \citenamefont {Yu},\ and\ \citenamefont
  {Agterberg}}]{Wu2024}%
  \BibitemOpen
  \bibfield  {author} {\bibinfo {author} {\bibfnamefont {H.}~\bibnamefont
  {Wu}}, \bibinfo {author} {\bibfnamefont {A.}~\bibnamefont {Amin}}, \bibinfo
  {author} {\bibfnamefont {Y.}~\bibnamefont {Yu}},\ and\ \bibinfo {author}
  {\bibfnamefont {D.~F.}\ \bibnamefont {Agterberg}},\ }\bibfield  {title}
  {\bibinfo {title} {{Nematic Bogoliubov Fermi surfaces from magnetic toroidal
  order in ${\mathrm{FeSe}}_{1\ensuremath{-}x}{\mathrm{S}}_{x}$}},\ }\href
  {https://doi.org/10.1103/PhysRevB.109.L220501} {\bibfield  {journal}
  {\bibinfo  {journal} {Phys. Rev. B}\ }\textbf {\bibinfo {volume} {109}},\
  \bibinfo {pages} {L220501} (\bibinfo {year} {2024})}\BibitemShut {NoStop}%
\bibitem [{\citenamefont {Mizukami}\ \emph {et~al.}(2023)\citenamefont
  {Mizukami}, \citenamefont {Haze}, \citenamefont {Tanaka}, \citenamefont
  {Matsuura}, \citenamefont {Sano}, \citenamefont {B{\"{o}}ker}, \citenamefont
  {Eremin}, \citenamefont {Kasahara}, \citenamefont {Matsuda},\ and\
  \citenamefont {Shibauchi}}]{Mizukami2023}%
  \BibitemOpen
  \bibfield  {author} {\bibinfo {author} {\bibfnamefont {Y.}~\bibnamefont
  {Mizukami}}, \bibinfo {author} {\bibfnamefont {M.}~\bibnamefont {Haze}},
  \bibinfo {author} {\bibfnamefont {O.}~\bibnamefont {Tanaka}}, \bibinfo
  {author} {\bibfnamefont {K.}~\bibnamefont {Matsuura}}, \bibinfo {author}
  {\bibfnamefont {D.}~\bibnamefont {Sano}}, \bibinfo {author} {\bibfnamefont
  {J.}~\bibnamefont {B{\"{o}}ker}}, \bibinfo {author} {\bibfnamefont
  {I.}~\bibnamefont {Eremin}}, \bibinfo {author} {\bibfnamefont
  {S.}~\bibnamefont {Kasahara}}, \bibinfo {author} {\bibfnamefont
  {Y.}~\bibnamefont {Matsuda}},\ and\ \bibinfo {author} {\bibfnamefont
  {T.}~\bibnamefont {Shibauchi}},\ }\bibfield  {title} {\bibinfo {title}
  {{Unusual crossover from Bardeen-Cooper-Schrieffer to
  Bose-Einstein-condensate superconductivity in iron chalcogenides}},\ }\href
  {https://doi.org/10.1038/s42005-023-01289-8} {\bibfield  {journal} {\bibinfo
  {journal} {Communications Physics}\ }\textbf {\bibinfo {volume} {6}},\
  \bibinfo {pages} {183} (\bibinfo {year} {2023})}\BibitemShut {NoStop}%
\bibitem [{\citenamefont {Islam}\ and\ \citenamefont
  {Chubukov}(2024)}]{Islam2024}%
  \BibitemOpen
  \bibfield  {author} {\bibinfo {author} {\bibfnamefont {K.~R.}\ \bibnamefont
  {Islam}}\ and\ \bibinfo {author} {\bibfnamefont {A.}~\bibnamefont
  {Chubukov}},\ }\bibfield  {title} {\bibinfo {title} {{Unconventional
  superconductivity near a nematic instability in a multi-orbital system}},\
  }\href {https://doi.org/10.1038/s41535-024-00638-2} {\bibfield  {journal}
  {\bibinfo  {journal} {npj Quantum Materials}\ }\textbf {\bibinfo {volume}
  {9}},\ \bibinfo {pages} {28} (\bibinfo {year} {2024})}\BibitemShut {NoStop}%
\bibitem [{\citenamefont {Bourgeois-Hope}\ \emph {et~al.}(2019)\citenamefont
  {Bourgeois-Hope}, \citenamefont {Li}, \citenamefont {Laliberté},
  \citenamefont {Badoux}, \citenamefont {Hayden}, \citenamefont {Momono},
  \citenamefont {Kurosawa}, \citenamefont {Yamada}, \citenamefont {Takagi},
  \citenamefont {Doiron-Leyraud},\ and\ \citenamefont
  {Taillefer}}]{BourgeoisHope2019}%
  \BibitemOpen
  \bibfield  {author} {\bibinfo {author} {\bibfnamefont {P.}~\bibnamefont
  {Bourgeois-Hope}}, \bibinfo {author} {\bibfnamefont {S.~Y.}\ \bibnamefont
  {Li}}, \bibinfo {author} {\bibfnamefont {F.}~\bibnamefont {Laliberté}},
  \bibinfo {author} {\bibfnamefont {S.}~\bibnamefont {Badoux}}, \bibinfo
  {author} {\bibfnamefont {S.~M.}\ \bibnamefont {Hayden}}, \bibinfo {author}
  {\bibfnamefont {N.}~\bibnamefont {Momono}}, \bibinfo {author} {\bibfnamefont
  {T.}~\bibnamefont {Kurosawa}}, \bibinfo {author} {\bibfnamefont
  {K.}~\bibnamefont {Yamada}}, \bibinfo {author} {\bibfnamefont
  {H.}~\bibnamefont {Takagi}}, \bibinfo {author} {\bibfnamefont
  {N.}~\bibnamefont {Doiron-Leyraud}},\ and\ \bibinfo {author} {\bibfnamefont
  {L.}~\bibnamefont {Taillefer}},\ }\bibfield  {title} {\bibinfo {title} {{Link
  between magnetism and resistivity upturn in cuprates: a thermal conductivity
  study of La$_{2-x}$Sr$_x$CuO$_4$}},\ }\href
  {https://arxiv.org/abs/1910.08126} {\bibfield  {journal} {\bibinfo  {journal}
  {arXiv:1910.08126}\ } (\bibinfo {year} {2019})}\BibitemShut {NoStop}%
\bibitem [{\citenamefont {Chareev}\ \emph {et~al.}(2013)\citenamefont
  {Chareev}, \citenamefont {Osadchii}, \citenamefont {Kuzmicheva},
  \citenamefont {Lin}, \citenamefont {Kuzmichev}, \citenamefont {Volkova},\
  and\ \citenamefont {Vasiliev}}]{Chareev2013}%
  \BibitemOpen
  \bibfield  {author} {\bibinfo {author} {\bibfnamefont {D.}~\bibnamefont
  {Chareev}}, \bibinfo {author} {\bibfnamefont {E.}~\bibnamefont {Osadchii}},
  \bibinfo {author} {\bibfnamefont {T.}~\bibnamefont {Kuzmicheva}}, \bibinfo
  {author} {\bibfnamefont {J.-Y.}\ \bibnamefont {Lin}}, \bibinfo {author}
  {\bibfnamefont {S.}~\bibnamefont {Kuzmichev}}, \bibinfo {author}
  {\bibfnamefont {O.}~\bibnamefont {Volkova}},\ and\ \bibinfo {author}
  {\bibfnamefont {A.}~\bibnamefont {Vasiliev}},\ }\bibfield  {title} {\bibinfo
  {title} {{Single crystal growth and characterization of tetragonal
  FeSe$_{1-x}$ superconductors}},\ }\href {https://doi.org/10.1039/c2ce26857d}
  {\bibfield  {journal} {\bibinfo  {journal} {CrystEngComm}\ }\textbf {\bibinfo
  {volume} {15}},\ \bibinfo {pages} {1989} (\bibinfo {year}
  {2013})}\BibitemShut {NoStop}%
\bibitem [{\citenamefont {Van~Degrift}(1975)}]{van1975tunnel}%
  \BibitemOpen
  \bibfield  {author} {\bibinfo {author} {\bibfnamefont {C.~T.}\ \bibnamefont
  {Van~Degrift}},\ }\bibfield  {title} {\bibinfo {title} {Tunnel diode
  oscillator for 0.001 ppm measurements at low temperatures},\ }\href
  {https://doi.org/10.1063/1.1134272} {\bibfield  {journal} {\bibinfo
  {journal} {Review of Scientific Instruments}\ }\textbf {\bibinfo {volume}
  {46}},\ \bibinfo {pages} {599} (\bibinfo {year} {1975})}\BibitemShut
  {NoStop}%
\bibitem [{\citenamefont {Vannette}\ \emph {et~al.}(2008)\citenamefont
  {Vannette}, \citenamefont {Sefat}, \citenamefont {Jia}, \citenamefont {Law},
  \citenamefont {Lapertot}, \citenamefont {Bud’ko}, \citenamefont {Canfield},
  \citenamefont {Schmalian},\ and\ \citenamefont {Prozorov}}]{VANNETTE2008354}%
  \BibitemOpen
  \bibfield  {author} {\bibinfo {author} {\bibfnamefont {M.}~\bibnamefont
  {Vannette}}, \bibinfo {author} {\bibfnamefont {A.}~\bibnamefont {Sefat}},
  \bibinfo {author} {\bibfnamefont {S.}~\bibnamefont {Jia}}, \bibinfo {author}
  {\bibfnamefont {S.}~\bibnamefont {Law}}, \bibinfo {author} {\bibfnamefont
  {G.}~\bibnamefont {Lapertot}}, \bibinfo {author} {\bibfnamefont
  {S.}~\bibnamefont {Bud’ko}}, \bibinfo {author} {\bibfnamefont
  {P.}~\bibnamefont {Canfield}}, \bibinfo {author} {\bibfnamefont
  {J.}~\bibnamefont {Schmalian}},\ and\ \bibinfo {author} {\bibfnamefont
  {R.}~\bibnamefont {Prozorov}},\ }\bibfield  {title} {\bibinfo {title}
  {Precise measurements of radio-frequency magnetic susceptibility in
  ferromagnetic and antiferromagnetic materials},\ }\href
  {https://doi.org/https://doi.org/10.1016/j.jmmm.2007.06.018} {\bibfield
  {journal} {\bibinfo  {journal} {Journal of Magnetism and Magnetic Materials}\
  }\textbf {\bibinfo {volume} {320}},\ \bibinfo {pages} {354} (\bibinfo {year}
  {2008})}\BibitemShut {NoStop}%
\bibitem [{\citenamefont {Ikeda}\ \emph
  {et~al.}(2018{\natexlab{b}})\citenamefont {Ikeda}, \citenamefont {Worasaran},
  \citenamefont {Palmstrom}, \citenamefont {Straquadine}, \citenamefont
  {Walmsley},\ and\ \citenamefont {Fisher}}]{Fisher2018}%
  \BibitemOpen
  \bibfield  {author} {\bibinfo {author} {\bibfnamefont {M.~S.}\ \bibnamefont
  {Ikeda}}, \bibinfo {author} {\bibfnamefont {T.}~\bibnamefont {Worasaran}},
  \bibinfo {author} {\bibfnamefont {J.~C.}\ \bibnamefont {Palmstrom}}, \bibinfo
  {author} {\bibfnamefont {J.~A.~W.}\ \bibnamefont {Straquadine}}, \bibinfo
  {author} {\bibfnamefont {P.}~\bibnamefont {Walmsley}},\ and\ \bibinfo
  {author} {\bibfnamefont {I.~R.}\ \bibnamefont {Fisher}},\ }\bibfield  {title}
  {\bibinfo {title} {Symmetric and antisymmetric strain as continuous tuning
  parameters for electronic nematic order},\ }\href
  {https://doi.org/10.1103/PhysRevB.98.245133} {\bibfield  {journal} {\bibinfo
  {journal} {Phys. Rev. B}\ }\textbf {\bibinfo {volume} {98}},\ \bibinfo
  {pages} {245133} (\bibinfo {year} {2018}{\natexlab{b}})}\BibitemShut
  {NoStop}%
\end{thebibliography}%

\end{document}